%% file: paper.tex
\documentclass[iop]{emulateapj}
\usepackage[T1]{fontenc}
\usepackage{microtype,amsmath}
\usepackage[colorlinks,urlcolor=blue,citecolor=blue,linkcolor=blue]{hyperref}

\slugcomment{ApJ in press}
\shorttitle{ASGARD I. Overview \& First Results}
\shortauthors{Williams \textit{et al.}}

\newcommand\citeeg[1]{\citep[e.g.,][]{#1}}
\newcommand\fnu[1]{\footnote{\url{#1}}}
\newcommand\cyg{\object{Cyg~X\mbox{-}3}}
\newcommand\acyg{Cyg~X\mbox{-}3}
\newcommand\kep{\textit{Kepler}}
\newcommand\sgra{\object{Sgr~A$\!$*}}
\newcommand\sqd{deg$^2$}
\newcommand\apx{\ensuremath{\sim}}
\newcommand\uv{$u$\mbox{-}$v$}
\newcommand\smn{\ensuremath{\langle S \rangle}}

\input figaliases.tex

\begin{document}

\title{ASGARD: A Large Survey for Slow Galactic Radio Transients. \\
  I. Overview and First Results
}
\author{
  Peter K.~G. Williams,
  Geoffrey C. Bower,
  Steve Croft,
  Garrett K. Keating,
  Casey J. Law,
  and Melvyn C. H. Wright
}
\affil{Department of Astronomy, B-20 Hearst Field Annex~\#~3411,
  University of California, Berkeley, CA 94720-3411, USA}
\email{pwilliams@astro.berkeley.edu}

\begin{abstract}
  Searches for slow radio transients and variables have generally
  focused on extragalactic populations, and the basic parameters of
  Galactic populations remain poorly characterized. We present a large
  3~GHz survey performed with the Allen Telescope Array (ATA) that
  aims to improve this situation: ASGARD, the ATA Survey of Galactic
  Radio Dynamism. ASGARD observations spanned 2~years with weekly
  visits to 23~\sqd\ in two fields in the Galactic Plane, totaling
  900~hr of integration time on science fields and making it
  significantly larger than previous efforts. The typical blind
  unresolved source detection limit was 10~mJy. We describe the
  observations and data analysis techniques in detail, demonstrating
  our ability to create accurate wide-field images while effectively
  modeling and subtracting large-scale radio emission, allowing
  standard transient-and-variability analysis techniques to be
  used. We present early results from the analysis of two pointings:
  one centered on the microquasar Cygnus~X\mbox{-}3 and one
  overlapping the \textit{Kepler} field of view ($\ell = 76\degr$, $b
  = +13.5\degr$). Our results include images, catalog statistics,
  completeness functions, variability measurements, and a transient
  search. Out of 134 sources detected in these pointings, the only
  compellingly variable one is Cygnus~X\mbox{-}3, and no transients
  are detected. We estimate number counts for potential Galactic radio
  transients and compare our current limits to previous work and our
  projection for the fully-analyzed ASGARD dataset.
\end{abstract}


\defcitealias{sgdbvdh+03}{S+03}
\newcommand\sbhw{\citetalias{sgdbvdh+03}}
\defcitealias{bck+10}{PiGSS--I}
\newcommand\pgsi{\citetalias{bck+10}}
\defcitealias{bwb+11}{PiGSS--II}
\newcommand\pgsii{\citetalias{bwb+11}}

\section{Introduction}
\label{s:intro}

The technological developments of the past few decades have led to an
explosion of interest in the astronomical time domain. There has been
a recent boom at radio wavelengths, where the survey capabilities of
new and upgraded facilities represent order-of-magnitude improvements
over their predecessors
\citeeg{thekarl,theapertif,thealfalfa,theaskap,themeerkat,theska}. Among
the motivators for the construction of these facilities are known or
strongly-supported classes of highly-variable slow extragalactic radio
emitters such as active galactic nuclei \citep{laa+11}, orphan
$\gamma$-ray burst afterglows \citep{fks+01,lowg02}, radio supernovae
\citep{wpms02,bmr+09,mbg+10}, and tidal disruption events
\citep{r88,bgm+11,bzp+12}. We follow other authors in defining slow
variables as those whose emission evolves on timescales $\gtrsim$1~s;
this approximately corresponds to those that emit via incoherent,
rather than coherent, processes, and are typically identified using
image-domain techniques. The relatively sparse prior exploration of
the dynamic radio sky additionally highlights it as an exciting
discovery space \citep{clm04,fb11,fko+12}.

Much work has recently gone into the characterization of the
population of highly-variable slow extragalactic sources. Archival
studies have used existing large-area surveys, usually observed at 1.4
or 4.9~GHz, to search for rare events
\citep{lowg02,bsb+07,bfs+11,bs11,of11,bmg+11errated,bmg+11,thwb11errated,thwb11,fko+12}. Followup
observations of candidate or confirmed highly-variable sources have
been used to characterize the detailed properties of individual
objects \citep{dvbwh04,gyop+06,mbg+10,bzp+12}. Finally, dedicated
surveys have been devoted to the systematic discovery of highly
variable radio sources. Several have been undertaken with the Allen
Telescope Array \citep[ATA;][]{theata}, one of the first radio
observatories explicitly designed to be an efficient survey
instrument, including the ATA Twenty-centimeter Survey
\citep{cba+10errated,cba+10,cbk+11} and the Pi~GHz Sky Survey
\citep[PiGSS--I and PiGSS--II hereafter; Croft et al., 2012, in
  prep.]{bck+10,bwb+11}. \citet{ofb+11} describe most of these
previous studies in greater depth. \citet{fko+12} summarize the state
of the field and find that reliable detection of these sources remains
both a challenge and an opportunity.

In contrast, there has been relatively little work to investigate the
population of highly-variable slow \textit{Galactic} radio
emitters. Source classes contributing to this population include X-ray
binaries \citep{wgj+95,mb75}, active stellar binaries
\citep{h76,bf77,eab+08}, cool dwarfs \citep{bbb+01,b02,hbl+07},
pre-main-sequence stars \citep{bpb+03,fmr08,shb08}, and flare stars
\citep{g02,jkw89,rwgr03}. Towards the Galactic Center (GC), several
intriguing sources of ambiguous nature have been discovered, including
the Galactic Center Transient at 1.36~GHz \citep{zrg+92},
\object{A1742$-$28} at 0.96~GHz \citep{dwb+76},
\object{\mbox{CXOGC}~J174540.0$-$290031} at a variety of frequencies
\citep{bryz+05}, \object{GCRT~J1745$-$3009} at 0.33~GHz
\citep{hlk+05,hlr+06}, and \object{GCRT~J1742$-$3001} at 0.235~GHz
\citep{hwl+09}. An overall increase in the prevalence of apparent
radio variability is expected towards the Galactic plane (GP), and the
GC in particular, due to interstellar scintillation
\citep[e.g.,][]{sfgp89,r90,gr92,gh00,lrm+08,of11}. Pulsars and several
other well-known classes of variable Galactic sources are not
discussed here because they do not fall into the ``slow'' category;
see \citet{clm04} and references therein.

For many years, the best available data on Galactic radio variability
came from the work of \citet{tg83} and \citet{gt86}, who used the NRAO
91-m transit telescope to repeatedly survey the GP at 5~GHz. Recent
work has begun to expand and update these
results. \citet{hlk+02,hlk+03,hlr+06} have periodically monitored the GC at
long wavelengths (0.3~GHz) with the VLA and GMRT, and as indicated
above they have discovered several unusual transient
sources. \citet{bhw+10} conducted an archival search for GP transients
at 5~GHz by comparing the VLA survey described in \citet{bwh+94} with
CORNISH, the ``Co-Ordinated Radio 'N' Infrared Survey for High-mass
star formation'' \citep{phd08}, discovering a population of highly
variable sources and analyzing their statistical properties. Most
recently, \citet{ofb+11} conducted a 5~GHz VLA variability survey at
low Galactic latitudes with rapid multiwavelength followup,
discovering a candidate transient source and measuring more
statistical properties of the population.

In this paper we present ASGARD: the ATA Survey of Galactic Radio
Dynamism. Its primary goals are to perform a deep search for Galactic
radio transients and to measure the variability properties of a wide
variety of Galactic radio sources on day-to-year timescales. To this
end, we repeatedly observed 24 pointings near the GP at 3~GHz with the
ATA, visiting most pointings on a $\lesssim$weekly cadence over the
course of \apx 2~yr and obtaining \apx 900~hr of integration on our
science pointings. The large field of view (FOV) of the ATA allows us
to cover a relatively large region on the sky, and our frequent visits
provide thorough sampling of variability on a range of timescales. The
compact configuration of the ATA also makes our observations sensitive
to static extended structures in the GP, such as nonthermal radio
filaments \citep{yzmc84,lyzc08}, \ion{H}{2} regions
\citep{bnk+03,nhr+06}, and supernova remnants \citep{g94}.

We proceed by describing the ASGARD observations (\S\ref{s:obs}) and
data processing (\S\ref{s:analysis}) in detail. This work concerns
itself with a subset of the whole dataset (\S\ref{s:subset}),
representing \apx 10\% of the expected usable observations, while
another portion of the dataset has already been described elsewhere
\citep{wtb+11}. We present first results (\S\ref{s:results}) derived
from analysis of this subset, including deep images, source catalogs,
a transient search, and variability statistics. Finally we discuss our
current conclusions and the prospects of the fully-analyzed survey
(\S\ref{s:conc}), populating a ``log-$N$/log-$S$'' plot for Galactic
radio transients (Figure~\ref{f:sadlimits}). % SKIPALIAS

\section{Observations}
\label{s:obs}

We used the ATA to monitor two fields: an area around the Galactic
center (GC) spanning $-4.5\degr < \ell < 8\degr$, $|b| < 2\degr$; and
\apx 5~deg$^2$ towards Cygnus including the highly radio-variable
microquasar \cyg\ \citep{gk72} and a portion of the FOV of the
\kep\ \citep{thekepler} mission. Coordinates of the ASGARD pointing
centers are listed in Table~\ref{t:pointings}, along with the
observing time devoted to each pointing over the course of the survey.
Throughout this work we use the word ``field'' to refer to either of
the two survey regions (GC and Cygnus) and the word ``pointing'' to
refer to a specific pointing center that was observed.

While some ASGARD observations were conducted with complete control
of the ATA, the vast majority were conducted commensally with a SETI
(Search for Extraterrestrial Intelligence) survey \citep{bata09}.
Both of these projects were designed to increase the likelihood of
discovering rare Galactic events by targeting regions with high source
densities, as well as to take advantage of the data stream provided by
the \kep\ mission, which motivated the basic choice of survey fields.
The division of the survey area into two fields was also a practical
choice given year-round observations, since the GC region is
seasonally difficult or impossible to observe from the ATA site
(latitude $+40.8$\degr). The Cygnus region, on the other hand, can be
observed year-round, also has a very high source density, and contains
the benchmark source \cyg. The observed GC field is as centered on
\sgra\ as possible given visibility constraints. The overall footprint
of each field was determined by the joint needs of the commensal SETI
search and ASGARD. The former required a certain dwell time per
pointing to sequentially survey targets using the ATA digital
beamformer backends. The latter aimed for a weekly revisitation
cadence. Combining these requirements with the typical weekly survey
time allocation determined the total footprint that could be
observed. The system used to organize and execute the commensal
observations is described in \citet{w12}.

The pointings in the GC field fall on an 11$\degr$$\times$2$\degr$ square grid in
Galactic coordinates with a spacing of 1\degr\ and a northeast corner
located at $\ell = 6.5\degr$, $b = 0.5\degr$. The half-power beam
width (HPBW) of the ATA is approximately $(3.5\textrm{ deg GHz}) /
\nu_\mathrm{obs}$ \citep[but see \S\ref{s:pb}]{hbc+10,han+11}, so that
the pointing centers range from slightly oversampled to critically
sampled for ASGARD observing frequencies of \apx 1--3~GHz. The Cygnus
field consists of an analogous 2$\times$2 grid, with \cyg\ located at
the northwest pointing center ($\ell = 79.8\degr$, $b = 0.7\degr$), as
well as a disjoint pointing toward $\ell = 75.8\degr$, $b = 13.5\degr$
($\alpha = 19^\textrm{h}21^\textrm{m}24^\textrm{s}$, $\delta =
44\degr00'00''$, ICRS J2000), which lies within the \kep\ mission
FOV. The total footprint of the survey on the sky is \apx 23~\sqd\ if
each pointing is taken to image a circle with a diameter of the
nominal HPBW.

At the time of the observations the ATA frontend consisted of
forty-two 6.1-m offset Gregorian dishes. The backends used for this
work were two FX correlators with bandwidths of 104.9~MHz divided into
1024 channels and a dump time of 10~s. The correlators accepted 64
``antpol'' inputs, meaning that they could perform full-Stokes
correlation of 32 antennas or, hypothetically, single-Stokes
correlation of 64 antennas. Because of the desirability of full-Stokes
coverage and an ongoing program of feed retrofits, the correlators
generally accepted data from \apx 32 distinct antennas. The set of
antennas used varied with time due to maintenance or hardware
failures, and the two correlators did not necessarily have identical
antpol inputs.

Each observing session (``epoch'') generally began with a long (\apx
30~min) observation of a bright, unresolved source, usually one of
3C\,48, 3C\,147, 3C\,286, or 3C\,295. These observations were
necessary for delay calibration of the beamformers used in the SETI
survey but also provided excellent bandpass and flux density scale
calibration data for ASGARD. Science pointings were observed with
periodic (every \apx 45~min) visits to a nearby phase
calibrator. Because of the low declinations of the GC pointings, these
were generally observed at relatively low elevations with constrained
hour angle ranges.

The ATA feeds have a high-bandwidth design using a log-periodic
architecture, where the location of the active region on the feed
varies with the observing frequency. Because the optics of the ATA
reflectors are frequency-independent, each feed is mounted on a piston
drive so that the appropriate part of the feed may be moved to the
optical focus for each observation. Focus positions are identified by
the optimal corresponding observing frequency, i.e., a focus position
of $\nu_\mathrm{foc} = 1.4$~GHz provides the best configuration for
observations at that frequency. The relationship between the piston
position and observing frequency is based on both theoretical and
empirical analysis \citep{han+11}. Defocused observations are possible
with some loss in system performance. For $\nu_\mathrm{foc} \gtrsim
0.9 \nu_\mathrm{obs}$, the penalty is slight, while for lower focus
frequencies (active region too close to the secondary) sensitivity
degrades and the primary beam (PB) broadens. For ASGARD observations
made in complete control of the array, the focus position was set
optimally, usually at 3.14~GHz. For commensal observations, the focus
position was set at 1.90~GHz.

The overall coverage statistics of ASGARD observations are recorded in
Table~\ref{t:campaigninfo}. We divide the observations into four
campaigns: three seasons of GC observations during the summers of
2009--2011, and regular Cygnus observations from 2009~November to
2011~April. For reasons discussed in the following section, we isolate
the set of observations made with the ATA correlators tuned to sky
frequencies of 3.04 and 3.14~GHz, which we refer to as the ``3~GHz''
observations. These are the same frequencies used by the PiGSS
survey. The overall amount of observatory time dedicated to the
project was 1650~hr, with 902~hr spent observing science targets.

\section{Analysis}
\label{s:analysis}

We mostly use standard techniques to calibrate and image the data,
using tools from the MIRIAD data reduction package \citep{themiriad}
for many steps, with several customized steps implemented in the
Python\fnu{http://python.org/} programming language via the package
\textsf{miriad-python} \citep{themirpy}. We use our images to
construct a catalog of every compact source detected in the survey and
to measure flux densities (or upper limits) for every observation of every
source. We perform a variability analysis on this photometric
dataset. We define a criterion by which sources may be classified as
``transient'' or not, but the classification does not affect the
variability analysis, and no transient sources are detected.

The chief difficulty in the particular case of ASGARD is successful
imaging of the significant large-scale structure (LSS) present in
virtually all ASGARD fields of view, with the notable exception of the
\kep\ pointing. Not only is LSS imaging generally challenging, but by
the nature of the survey most ASGARD epochs have sparse hour angle
coverage. Fortunately, the LSS emission is expected to be
time-invariant, while any astrophysical transients will be unresolved
by the ATA (resolution \apx 1$'$ at 3~GHz). All observations of a
given pointing can therefore be combined to produce an LSS model that
can then be subtracted in the visibility domain from each epoch's
observations. To ease the measurement of the variability of compact
sources, we exclude these sources from the LSS model, causing them to
remain in the per-epoch images. Putting aside for the moment the
important issue of subtraction errors, this approach yields per-epoch
images consisting only of compact sources upon which standard
transient-search techniques may be used. We describe our
implementation of this general approach below.

\subsection{Subset of Data Presented in This Work}
\label{s:subset}

The first season of GC observations was performed at sky frequencies
of 1.43 and 2.01~GHz. It was found, however, that there was
substantial broadband interference at these frequencies that presented
numerous challenges for data analysis \citep{w10}. Subsequent
observations primarily used the PiGSS 3~GHz setup, although a few
observations were made at other frequencies. The analysis presented in
this work is restricted to 3~GHz observations. In
Tables~\ref{t:pointings} and \ref{t:campaigninfo} we provide coverage
statistics as computed for the 3~GHz subset of the complete ASGARD
dataset.

In this work, we analyze two particular ASGARD pointings. The first
is towards the highly-variable source \cyg, which allows us to
demonstrate the detection of variable radio sources embedded in
complex, large-scale emission. The second is the \kep\ pointing, which
allows us to investigate the performance of our techniques in a field
without unusual imaging challenges. We demonstrate the analysis of 29
epochs of the \cyg\ pointing, representing 34\% of the 3~GHz
\cyg\ dataset by number of epochs and 17\% by raw data volume, and six
epochs of the \kep\ pointing, five of which also involved visits to
the \cyg\ pointing. These epochs nearly completely sample the timespan
of the 3~GHz Cygnus campaign, ranging from 2010~February~03 to
2011~April~04, and are roughly uniformly spaced in time. We have
processed and imaged epochs surrounding the \cyg\ radio flares of
2010~May \citep{bss+10} and 2011~March \citep{kmt+11}, but do not run
our full pipeline on the latter epochs, of which there are six,
because the 20~Jy flare leads to severe imaging problems related to
dynamic range limitations. Tests of our transient detection process
confirm, however, that it succeeds in this trivial case. Other work
describes the ATA-42 observations of these events in more detail
\citep{wtb+11}. A somewhat larger portion of our observations has been
processed and analyzed but without significant multi-epoch coverage of
other pointings, and so is not presented in this work to maintain a
clear focus on the better-covered \cyg\ and \kep\ pointings.

\subsection{Calibration \& Flagging}

Radiofrequency interference (RFI) was an intermittent problem during
the 3~GHz observations. The forms of RFI most commonly encountered
were narrow-bandwidth (1--5 channel) tones that affected the majority
of baselines with a \apx 100\% duty cycle, wider-bandwidth tones (\apx
5~MHz) that affected moderate numbers of baselines with a moderate
duty cycle, and brief ($<1$ dump) broadband bursts affecting all
baselines. Problems in the digital hardware (e.g., overheating) could
also cause RFI-like effects, typically manifesting themselves as
complete corruption of one half or one quarter of the spectrum for
certain baselines. In the results we report, RFI was primarily excised
from the data manually, using standard MIRIAD tools and an
interactive, graphical visibility visualizer for the RFI with more
complex time/frequency structure \citep{themirpy}. Several approaches
to automatic flagging have also been pursued with ATA data
\citeeg{kbw10,bck+10} and their integration into the ASGARD pipeline
is being investigated.

Standard bandpass and gain calibration techniques are used. The long
calibration observations at the beginning of each epoch are used to
set the flux density scale, referencing to \citet{bgpt+77}. For the GC
observations, the gain calibrator was \object{NRAO~530} ($\ell =
12.03, b = +10.81$), while for the \cyg\ observations it was usually
\object{BL~Lac}. Because this latter source is variable and usually
\apx 10\% linearly polarized, gain parameters for it are derived from
the bandpass observations whenever possible, treating X and Y feeds
separately. Our observing program did not allow for planned
observations of polarimetric calibrators over wide hour angle ranges,
so we are unable to solve for the frequency-dependent leakage terms
that would be required for polarimetric calibration of the ATA
\citep{lgb+11} on an epoch-by-epoch basis.

\subsection{Imaging and Source Extraction}
\label{s:imaging}

After calibration, the data are averaged down to 16 spectral channels
of 6.5536~MHz bandwidth per correlator and are converted to CASA
\citep{thecasa} format for imaging. This is necessary because MIRIAD
does not implement any wide-field imaging algorithms, which we have
found to be necessary for our analysis. In particular, without the use
of techniques such as polyhedral imaging \citep{cp92} or
$w$-projection \citep{cgb08}, sources far away ($\gtrsim$0.8\degr)
from phase center do not deconvolve well and acquire an hour angle
dependence in their position, which is severely problematic for both
LSS modeling and photometric extraction. In the steps we describe
below, the averaged channels are gridded using multifrequency
synthesis \citep{themfs} and imaged using $w$-projection with 128
planes (but not polyhedral techniques). Images are 2048$\times$2048
with a pixel size of 10$''$ and thus span approximately five times the
HPBW at 3~GHz.

We first construct a deep image for each pointing. Because these
images are used to generate several important data products, the
imaging techniques vary from pointing to pointing depending on what
produces the best results empirically. For the \kep\ pointing, we use
the Cotton-Schwab deconvolution method \citep{s84,cbb99}, while for
\cyg\ we use the maximum-entropy algorithm \citep{themaxen} with a
Gaussian prior. To achieve consistent LSS sampling without requiring
detailed primary beam models, the data contributing to each deep image
have a consistent ATA feed focus position. Imaging artifacts, rather
than thermal noise, currently limit the image quality, so the use of
only part of our data to form the deep images does not significantly
affect their sensitivities.

The deep images are used to construct a catalog of compact sources,
the properties of which are described in \S\ref{s:catalog}. Sources
are detected using a combination of the MIRIAD task \textsf{sfind} and
manual inspection to check for missed detections and reject dubious
ones. (Given the resolution, sensitivity, and footprint of ASGARD, these
techniques are scalable to the whole survey.) Source positions,
shapes, and mean flux densities are cataloged. Because the primary aim of
ASGARD is to study variability, our catalog does not initially include
sources that are marginally detected in the deep images, since such
sources will typically not be detectable in the epoch images. The
epoch images, however, are searched for uncataloged sources as
described below, so that any variable source that becomes detectable
during an epoch will eventually be included in the catalog.

Each ASGARD pointing is also associated with a static LSS model. In
the \kep\ pointing, this model is blank; for the rest, the model is
derived from the deep image and the compact source catalog, using
either deconvolution of a variant deep image in which the compact
sources have been subtracted, or source-fitting techniques on the
model image. The latter approach can be helpful because
maximum-entropy deconvolution tends to model unresolved sources as
Gaussians about the size of the synthesized beam. Because there are
substantial numbers of \cyg\ observations made at focus positions of
both 1.90 and 3.14~GHz, we generate one LSS model of this pointing for
each focus position, so that the approximations of our primary beam
modeling scheme (\S\ref{s:pb}) do not lead to avoidable LSS
subtraction errors.

Images from individual epochs are made the same way as the deep
images, except that before imaging the appropriate LSS model is
subtracted from the \uv\ data, and baselines shorter than 50~m are not
imaged. Deconvolution of the individual epoch images is performed with
800 iterations of CASA's ``wide-field'' implementation of the H\"ogbom
CLEAN algorithm \citep{h74}. The restoring beam is chosen
automatically by the imaging software and has a typical size of $80''
\times 40''$. In two epochs of \cyg\ observations, there are small but
discernable errors in the flux density scale that lead to noticeable
residuals in the LSS subtraction. We fixed the scales in these epochs
by trial-and-error, adjusting and reimaging the data and visually
assessing the magnitude of the LSS subtraction residuals. The
correction factors are -1\% and 5\%, and we estimate that they are
uncertain by about 1\%. (Because the LSS is strong and unvarying, it
is possible to be relatively precise.) We discuss our investigations
into rigorous, global cross-calibration of the flux density scale in
\S\ref{s:photom}.

Lightcurves of the cataloged sources are derived using image-domain
fitting on the LSS-subtracted individual epoch images. Each fit holds
the source position and shape fixed but allows the total flux density
to vary. Very close sources are fit simultaneously. Because the
reality, positions and shapes of these sources are known from the deep
image, we use a fairly weak constraint and consider a source to be
detected if its fitted flux is more than three times the local
background rms. Undetected sources are cataloged with an upper limit
of this theshold.

During the source fitting process, a residual image is generated by
subtracting off the best-fit image-domain model of every detected
source. There are generally subtraction residuals around each source
of peak magnitude $\lesssim$10\%\ of the source flux, with both
positive and negative components. The mean residual is another \apx
10\%\ smaller because the fitting process tends to minimize this
value. Allowing the source shapes and/or positions to vary yields
smaller residuals by definition, but the derived parameter values vary
much more from epoch to epoch than instrumental and observational
uncertainties lead us to expect. We instead attribute the subtraction
residuals primarily to \uv\ calibration errors. In this case our
choice to fix the source positions and shapes gives a more realistic
assessment of flux uncertainties by avoiding overfitting of the data.

\subsection{Detection of Uncataloged Sources}
\label{s:sfind}

We use \textsf{sfind} to search for any residual sources in the
individual epoch images after subtraction of both the LSS and the
cataloged compact sources. The LSS is subtracted in the \uv\ domain but
the compact sources are subtracted in the image domain. In our
analysis, a ``transient'' is any source detectable in an individual
epoch image that is not detectable in the corresponding deep
image. Sources discovered via \textsf{sfind} are added to the ASGARD
catalog and so are subsequently processed in the same way as all
others. For those pointings in which the LSS model is derived from a
deep image, such a source's contribution to the deep image propagates
into the LSS model. This mean flux density is equal to the
bright-epoch flux density
diluted by \apx $1/n$, where $n$ is the number of epochs. This
contribution is subtracted from the per-epoch images used for residual
source detection but will not significantly alter detectability for
$n$ more than a few.

Previous searches for radio transients have typically been dominated
by false positives \citep{fko+12}. Our procedure involves multiple
rounds of sky modeling and subtraction which will inevitably leave
artifacts as well. Our transient-detection step therefore has
stringent detection limits and cross-checks for systematic effects. We
use the false-discovery rate (FDR) algorithm in \textsf{sfind}
\citep{thesfind}. The background rms is computed in 64$\times$64-pixel
boxes (\textsf{sfind} keyword \textsf{rmsbox}) and the target FDR is
set to 0.5\% (\textsf{sfind} keyword \textsf{alpha}). To assist in
computing detection limits, \textsf{sfind} was modified to report an
estimated minimum detectable source flux density in the event that no
sources were found in an image, basing this value on an estimated FDR
``$p$-value'' that would be needed to have yielded a source detection.
The source shapes reported by \textsf{sfind} are deconvolved from the
synthesized beam. (Recall that a typical synthesized beam size in our
images is $80'' \times 40''$.) Sources smaller than the synthesized
beam have their shape fixed to that of the beam and have their
parameters refit (\textsf{sfind} option \textsf{psfsize}). Those that
are otherwise incompatible with the synthesized beam shape (e.g., both
elongated with perpendicular major axes) are also treated as point
sources. This is a conservative approach because, as described in the
next paragraph, we reject extended sources.

Sources reported by \textsf{sfind} are filtered according to several
criteria. Sources for which \textsf{sfind} does not report a
positional uncertainty, usually indicative of a very poor fit, are
rejected. Because genuine astrophysical transients will not be
resolved by the ATA, sources in
which the product of the deconvolved major and minor axes exceeds
7000~arcsec$^2$ are rejected, as are those in which the deconvolved
major axis exceeds 130~arcsec. Sources for which the
modeled primary beam attenuation exceeds 98.9\% (separation of
$3\sigma_\mathrm{PB}$, where $\sigma_\mathrm{PB} = \mathrm{HPBW} /
\sqrt{8 \log 2}$) are rejected, although our later analysis uses much
more conservative PB attenuation cutoffs. Finally, newly-detected
sources near previously cataloged steady sources are also
rejected. The match radius for this test is $50f$~arcsec where $f$
depends on the cataloged total flux density $S_k$ of the known
source. For $S_k < 20$~mJy, $f = 1$; for $S_k > 2.97$~Jy, $f = 10$;
and for intermediate values, $f = 2 \log (S_k / 20\textrm{ mJy})$. All
of the above cutoff values were determined by examining the properties
of the transient candidates that were both significantly detected and
obviously spurious. Remaining candidates are examined manually as
described in \S\ref{s:transients}.

\subsection{Primary Beam Modeling}
\label{s:pb}

In order to compute accurate flux densities, sensitivities, and sky models, we
must account for the primary beam of the ATA. \citet{hbc+10} analyzed
the primary beam of the ATA using data from the PiGSS survey, finding
$\mathrm{HPBW} = 1.10 \pm 0.01\degr$ for a circular Gaussian primary
beam model. Because many of our observations are performed with the
same frequency configuration as PiGSS, we adopt this value assuming a
$\nu^{-1}$ frequency dependence, i.e. $\mathrm{HPBW}_\mathrm{optimal}
= (3.40\textrm{ deg GHz}) / \nu_\mathrm{obs}$ for a mean PiGSS observing
frequency of 3.09~GHz. The numerator is slightly smaller than the more
generic value of 3.5~deg~GHz reported by \citet{han+11}.

Our PB modeling is complicated, however, by the fact that a
substantial number of 3~GHz observations are made with the focus set
to 1.90~GHz. Although we do not have observations specifically aimed
at characterizing the PB shape of 3~GHz observations at this focus
setting, we measure this value from the data in two ways. Firstly, our
observations of the \cyg\ pointing have extensive coverage in both
this focus position and in the optimally focused position. By
comparing compact source flux densities in deep images made from the two sets
of data, we obtain $\mathrm{HPBW}_\mathrm{1900} = 3.92\textrm{ deg
  GHz} / \nu_\mathrm{obs}$. We also compare the ATA-apparent flux densities of
point sources observed in the \kep\ pointing to those from the NVSS
catalog (see \S\ref{s:catalog}), assuming a typical source spectral
index $\alpha$ of -0.7 ($S_\nu \propto \nu^\alpha$), and find a factor
of 3.96~deg~GHz. We use the former value. Below, we denote the PB
correction factor as applied to a particular location $f_\mathrm{PB}$,
defined such that $f_\mathrm{PB} \ge 1$.

Holographic measurements of the ATA dishes suggest that the PB pattern
becomes increasingly noncircular as the focus moves away from its
optimal setting \citep{han+11}. Due to alt-az mount of the ATA dishes,
the PB rotates on the sky over the course of an observation. Proper
accounting for such an effect would need to occur during the Fourier
inversion process, which neither MIRIAD nor the standard CASA imager
are capable of doing. We are thus unable to measure, or compensate
for, this effect. (PB rotation can be dealt with approximately by
imaging the data in blocks of similar hour angle, but this reduces
\uv\ coverage and thus hampers the deconvolution of our complex
fields.) The measurements of \citet{han+11} suggest that in our
configuration the axial ratio should be about 10\%.

We also assume that each dish has an identical PB pattern and that all
dishes are pointed identically. These assumptions allow us to handle
PB correction in the image domain and are relied on in some of the
analysis that follows. The ability to relax these assumptions, like
the ability to model PB rotation, relies on either painstaking
iterative approaches \citep{wc08} or support in imaging software that
is not yet widespread; two notable implementations are the MeqTrees
system \citep{ns10} and a derivative of the CASA imager equipped with
the $A$-projection algorithm \citep{bcgu08}. Adapting our analysis to
a pipeline in which PB correction occurred during the imaging process
would require actual measurement, rather than calculation, of the
spatial variation in image noise, but would not require major
conceptual changes.

\subsection{Multi-Epoch Photometry}
\label{s:photom}

Our imaging and source extraction pipeline yields multi-epoch
photometry for our catalog sources, though for some faint sources our
measurements yield mostly upper limits. Each source additionally has a
flux density measurement from the deep image in which it is best
detected. Some bright sources are detectable in the deep images of
multiple pointings.

Our images are not corrected for the attenuation of the ATA primary
beam, so we must correct our flux density measurements for this effect. In the
subset of data we consider, all \kep\ observations are made with a
focus setting of 1.90~GHz, and so the primary beam correction for a
given source is the same in every epoch. The \cyg\ observations are
made with focus settings of both 1.90 and 3.14~GHz, so the primary
beam correction can vary from epoch to epoch. The deep \cyg\ image
derives from the optimally-focused data. Comparison of the
\cyg\ flux densities is complicated by the fact that we use different LSS
models for each focus setting. While this is necessary to accurately
subtract LSS from each epoch as well as possible, differences in the
LSS model around each sourch can lead to an additive flux density offset
beween measurements made at different focus settings, above and beyond
any multiplicative errors caused by the limitations of our analytic
primary beam models. In this work, we choose to consider only
measurements from a consistent focus setting, choosing the one that
resulted in the higher mean detection significance. For sources close
to the pointing center, these are generally the optimally-focused
data, whereas for sources far from the pointing center, the broad
defocused primary beam results in more significant detections.

Some images yield particularly poor photometric results and we discard
these measurements. These are generally either epochs with very few
contributing data or those from 2011~March in which \cyg\ was
undergoing a major (\apx 20~Jy) flare, leading to significant dynamic
range issues. Even without the inclusion of the latter flare in our
analysis, \cyg\ is the most significantly variable source in our
dataset.

We set the uncertainty on each flux density measurement to be
\begin{equation}
\sigma^2 = \sigma_\mathrm{rms}^2 + (0.05 S)^2 +
(0.5\mathrm{~mJy})^2,
\label{e:uncert}
\end{equation}
where $\sigma_\mathrm{rms}$ is the background rms of the relevant
image-domain fit and $S$ is the measured value. The constants in this
equation, which are in the range typically encountered in radio
interferometry, were chosen to yield a plausible distribution of $P_c$
values (see \S\ref{s:var}) for the least-variable survey sources. We
do not include the uncertainty of the flux density scale correction
factor for the two epochs that were rescaled (\S\ref{s:imaging}) or
attempt to quantify the uncertainty on the PB correction. We
investigated a correction for CLEAN bias \citep{thefirst,ccg+98} but
did not find compelling evidence that this improved our results, so we
do not include such a correction in our analysis. We also investigated
but did not use parabolic photometric models as employed for some
sources by \citet{bmg+11errated,bmg+11}. The full ASGARD dataset
should have enough flux density measurements to be able to investigate
the impact of these and other techniques in a more statistically
rigorous manner.

\citet{bmg+11errated,bmg+11} and \citet{ofb+11} used ``post-imaging
calibration'' (PIC) techniques in which they cross-calibrated their
photometry assuming that most sources do not vary systematically and
that each image is subject to a multiplicative flux density scale
correction. The flux density scale correction that we describe in
\S\ref{s:imaging} contrasts in that it happens before the imaging
process, is only applied to two problematic epochs, and uses
calibration factors determined in a clearly more ad-hoc way. We
investigated but did not use a ``pre-imaging calibration,'' in which
we correct the flux density scale of every epoch's visibilities using
a correction factors derived from the photometric data as in
PIC. Besides being extremely computationally costly, since applying
the calibration requires reimaging the entire survey, we found in
small-scale tests that the calibration factors were not stable over
multiple iterations of the calibration. We also found that this
approach did not correct the LSS subtraction residuals in the two
problematic epochs nearly as well as we could manually. We
additionally considered but did not use a true PIC, operating only in
the photometric domain. As was also found in \pgsii, we found that
such a calibration sometimes had encouraging results, especially for
bright sources, but that its effect on the photometry of faint sources
appeared to range between neutral and negative. As such we decided to
err on the side of simplicity and fewer data transformations.

For those compact sources that are found within a region of more
extended emission, our LSS subtraction technique naturally leads to
the possibility of an additive flux density bias in our measurements. We
constructed a set of images in which the LSS model was added back to
each epoch image after convolution with the synthesized beam. We then
extracted photometry again using a specialized routine that
simultaneously fit for the flux density of a source in each image on top of a
local, constant background term. This more elaborate procedure did not
produce more consistent results, and we found that our LSS subtraction
procedure did not lead to consistent flux density biases.

\begin{figure}[p]
\plotone{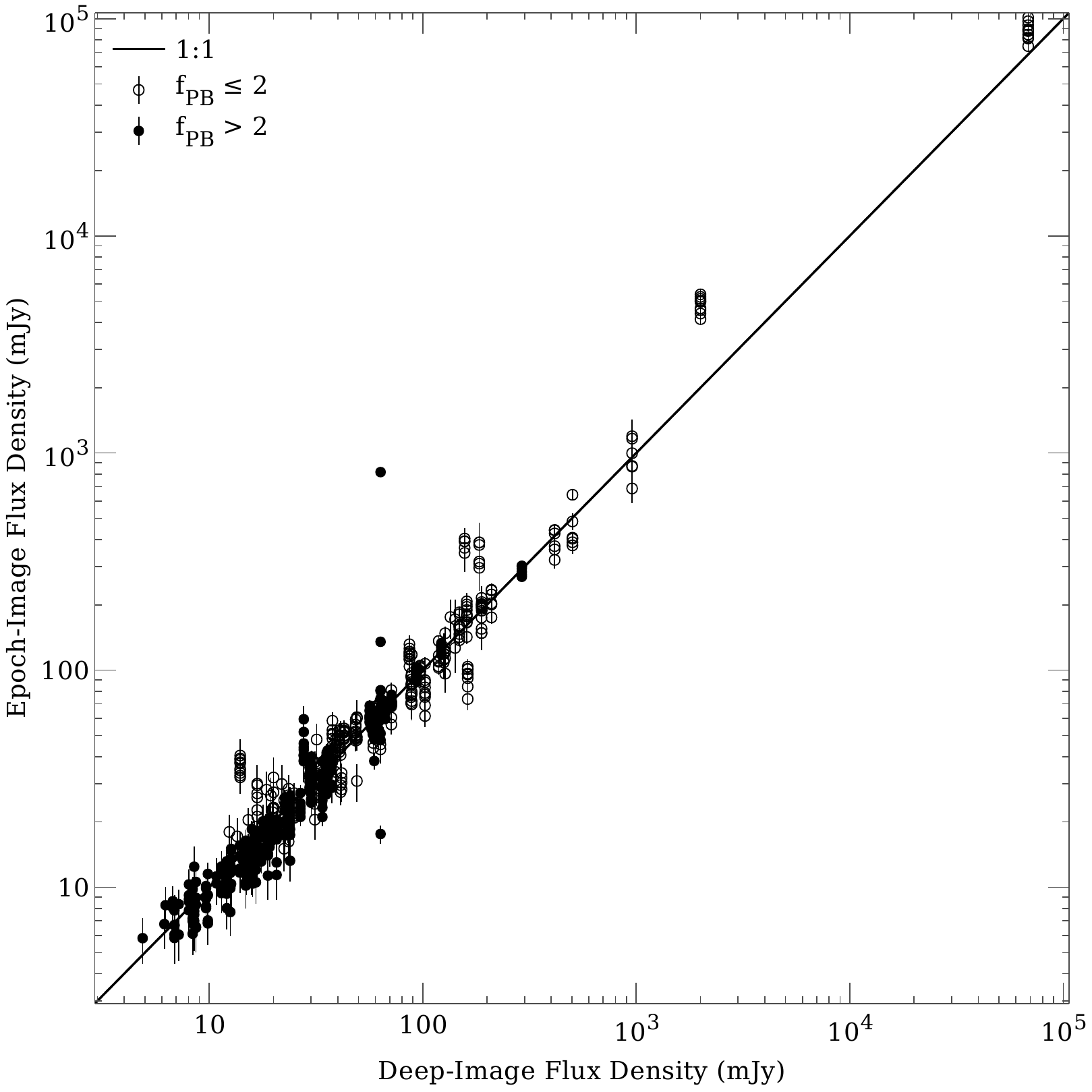}
\caption{Comparison between flux densities of cataloged sources as measured in
  the deep images and those measured in individual epochs. Agreement
  is generally very good, especially for those sources for which the
  primary beam correction is not large. The vertical set of points
  with a large spread corresponds to the highly variable source
  \acyg. Several sources have a discernable offset between the deep
  image and epoch flux densities; these are seen in \acyg\ pointing, and the
  offset arises from disagreements between the large-radius behavior
  of the optimally focused PB model (used to obtain the deep flux density) and
  that of the defocused PB model (used to open the epoch flux densities). The
  nominally very bright sources are subject to large, uncertain PB
  corrections; see \S\ref{s:photom}.}
\label{f:epochvsdeep}
\end{figure}

In Figure~\ref{f:epochvsdeep} we compare our deep and epoch flux
density measurements after correcting the flux density scale of the
two epochs that needed it, accounting for PB attenuation, discarding
low-quality points, and augmenting the uncertainties as per
Equation~\ref{e:uncert}. There is very good agreement in the scales
between the two. \cyg\ is clearly the most significantly variable
source we observe. The nominally very bright sources are all detected
at very large distances from the pointing center and are subject to
highly uncertain primary beam corrections. In order of decreasing deep
image flux density, these are \object{DR\,22} (separation 110$'$; PB
correction \apx 340), \object{DR\,21} (80$'$; \apx 25), and
\object{4C\,44.32} (99$'$; \apx 108). The last factor listed obeys a
different Gaussian relation than the first two because of the
different focus settings used in the \kep\ and \cyg\ deep images.

\subsection{Timescales}
\label{s:timescales}

\begin{figure}
\plotone{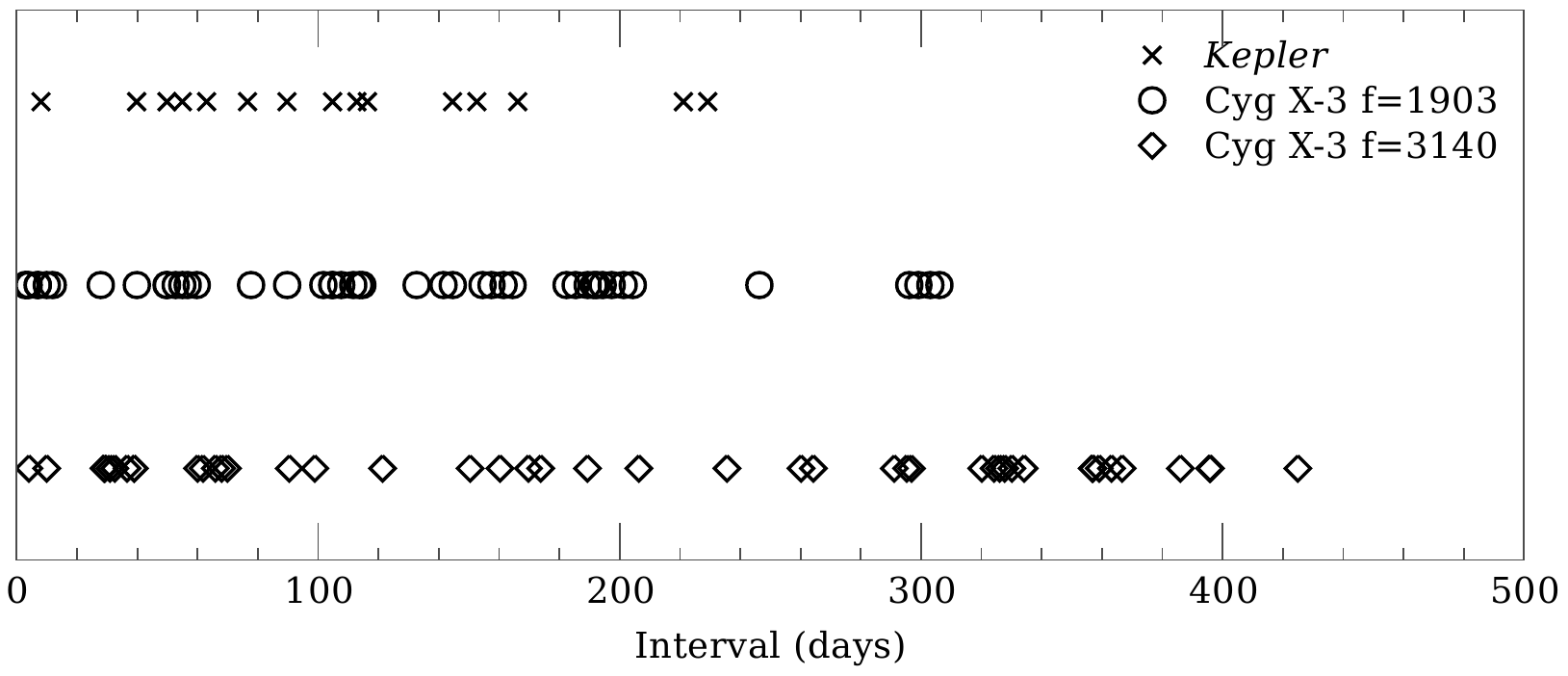}
\caption{Intervals between observations for the three sampling
  patterns present in this work. Different samples are not independent
  because for $n$ samples there are $n(n-1)/2$ intervals probed. A
  higher density of measurements around a certain timescale, however,
  generally indicates increased sensitivity to variability on that
  timescale.}
\label{f:intervals}
\end{figure}

We sample the variability of our catalog sources on timescales of days
to years. If a given source is measured $n$ times at dates $t_i$,
there are $n(n-1)/2$ possibly-redundant intervals sampled, $|t_i -
t_j|$ for $i < j$. We plot these intervals in Figure~\ref{f:intervals}
for the three samplings in this work. Coverage is fairly uniform
across the range of probed timescales.

With additional analysis, our observations can also be sensitive to
variability on timescales around the observation time of a typical
epoch image, \apx 1~hour. Evolution at and below this threshold can be
searched for by imaging subsets of the data within each epoch, with
the expected tradeoff between time resolution and sensitivity. Our
maximum time resolution is the ATA correlator integration time of
10~s. Any single-epoch transients discovered in the complete dataset
will be examined for intra-epoch evolution in this manner, but in this
work we find no sources that merit this detailed investigation. In the
gap between our \apx hour integration time and our \apx week observing
cadence, we are potentially sensitive to events (e.g., a lucky
observation may occur precisely during an hour-long stellar flare) but
can only poorly constrain their duration.

Our sensitivity to slowly-evolving sources might correspondingly be
improved by imaging, e.g., months of data at a time, and searching for
sources too faint to be detected in a single epoch but too variable to
be detected in the deep image, as in \citet{bsb+07} and other
works. (This is simply a rough matched-filter approach.) Because our
images are not limited by thermal noise it is unclear how much of a
benefit this technique would provide in practice. We defer an
investigation of this approach to future work.

\section{First Results}
\label{s:results}

In the following subsections we present the first results from
analyzing the subset of ASGARD data described above.

\subsection{Images}
\label{s:images}

\begin{figure}[p]
\plotone{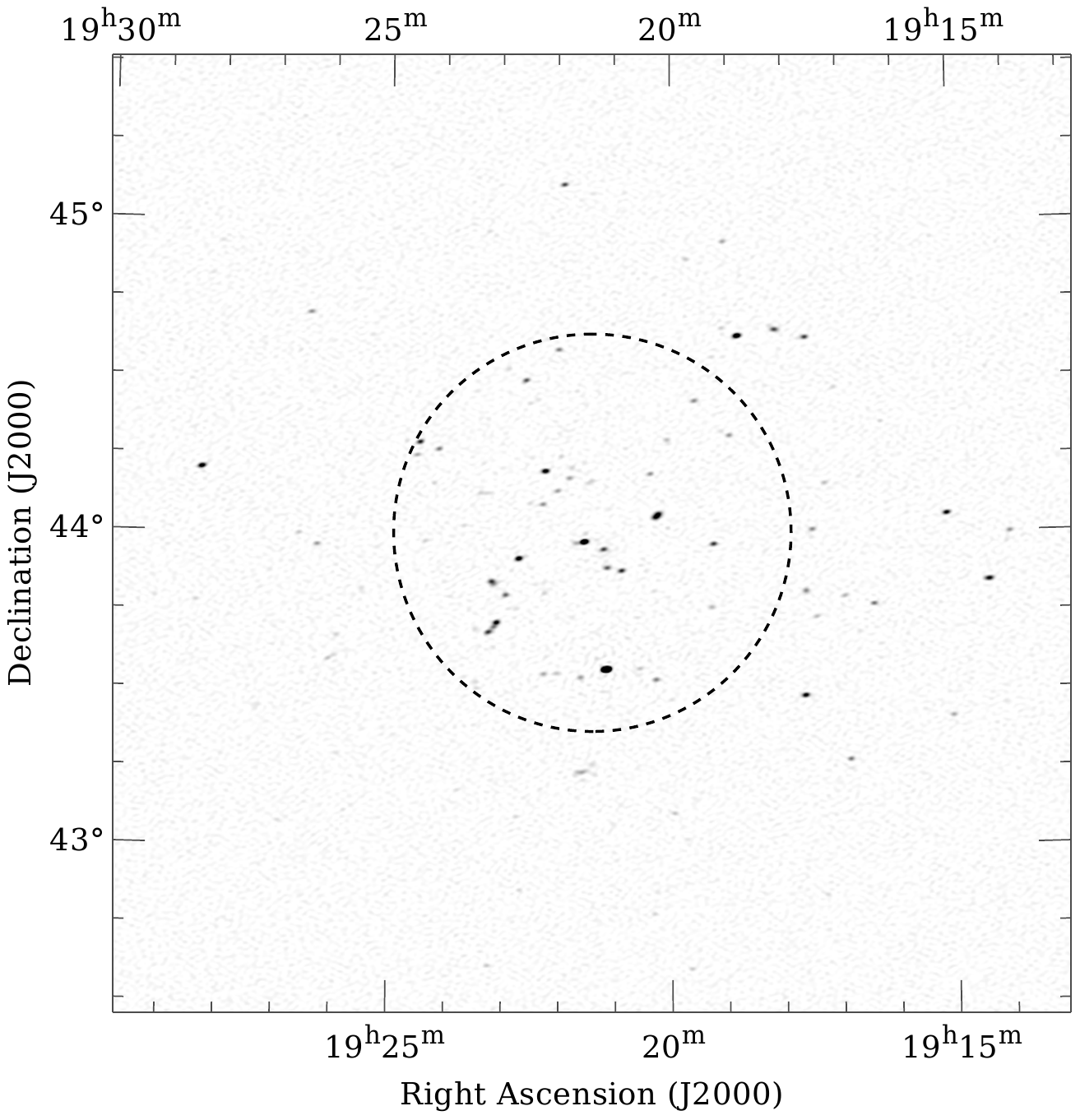}
\caption{Deep image of the \textit{Kepler} pointing, uncorrected for
  primary beam attenuation. The grayscale is linear from zero (white)
  to 20~mJy/beam (black). The maximum brightness in the image is \apx
  200~mJy/beam. The rms residual after applying the CLEAN algorithm is
  0.5~mJy/beam. The HPBW for this image is 1.27\degr\ and is denoted
  by the dashed circle. The synthesized beam is $65'' \times 41''$ at
  a position angle of $98\degr$.}
\label{f:kepdeep}
\end{figure}

The deep image of the \kep\ pointing at 3~GHz is shown in
Figure~\ref{f:kepdeep}. All of the observations of this pointing were
made with a focus setting of 1.9~GHz, so the primary beam is broadened
as compared to the nominal, optimally-focused configuration.

\begin{figure}[p]
\plotone{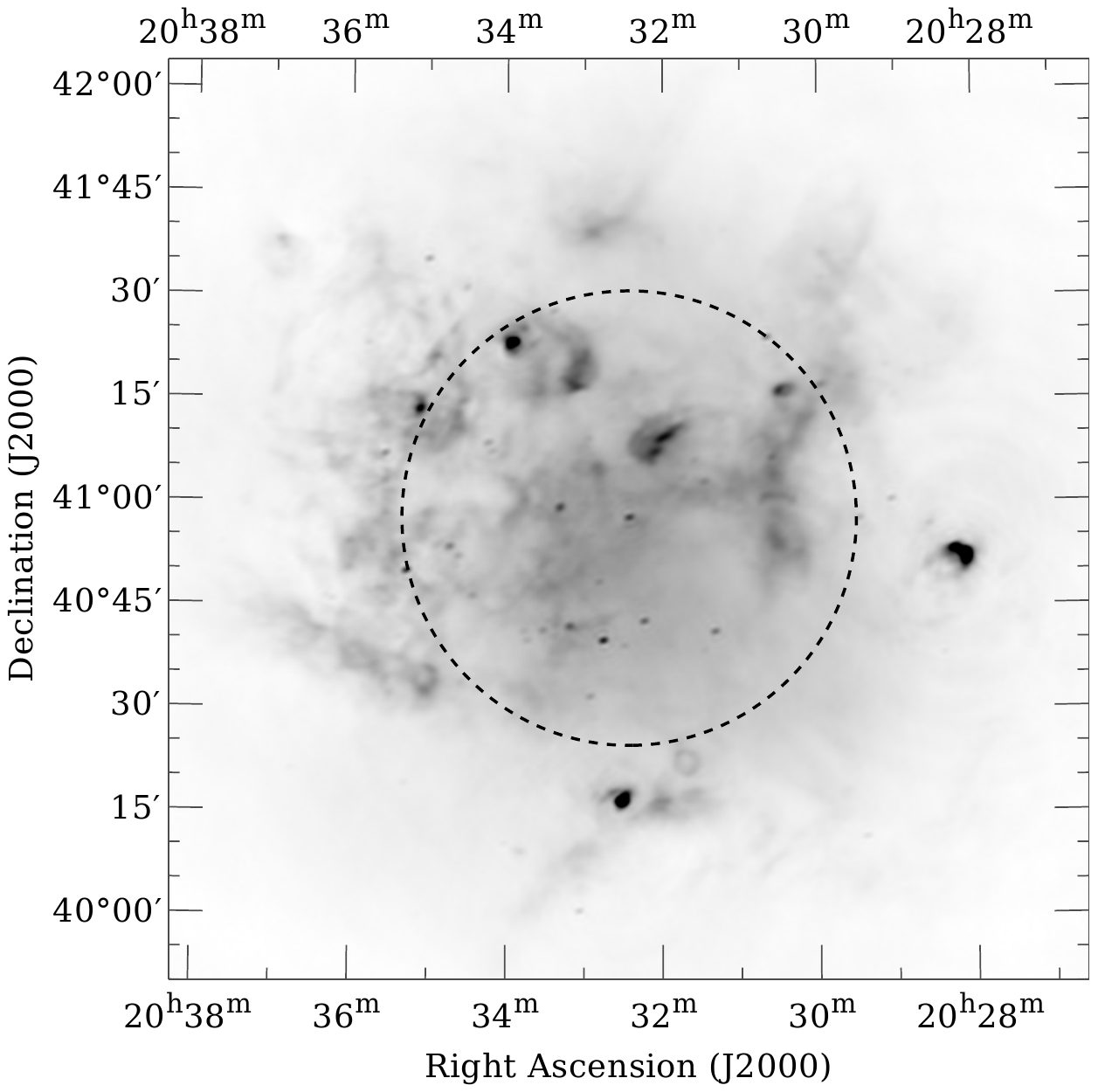}
\caption{Deep image of the \acyg\ pointing, uncorrected for primary
  beam attenuation. The grayscale is linear from zero (white) to
  150~mJy/beam (black). The maximum brightness in the image is \apx
  500~mJy/beam. The rms deconvolution residual is 1~mJy/beam. A
  Gaussian prior image was used in the maximum-entropy deconvolution
  process, causing the restored image to contain emission on angular
  scales larger than those sampled by the ATA data. The HPBW for this
  image is 1.1\degr\ and is denoted by the dashed circle. The
  synthesized beam is $60'' \times 42''$ at a position angle of
  $106\degr$.}
\label{f:deep}
\end{figure}

The optimally-focused deep image of the \cyg\ pointing at 3~GHz is
shown in Figure~\ref{f:deep}. The LSS and FOV may be compared with the
1.4~GHz image made with the Westerbork Synthesis Radio Telescope
(WSRT) presented in \citet[fig.~2; S+03 hereafter]{sgdbvdh+03}. The
ATA's compact configuration and large FOV make it much more sensitive
to LSS than the WSRT, and it is notable that the image shown in
Figure~\ref{f:deep} is produced from (multiple epochs of) a single
pointing. We detect significant LSS out to a radius of \apx 45$'$ from
the pointing center.

\begin{figure}[p]
\plotone{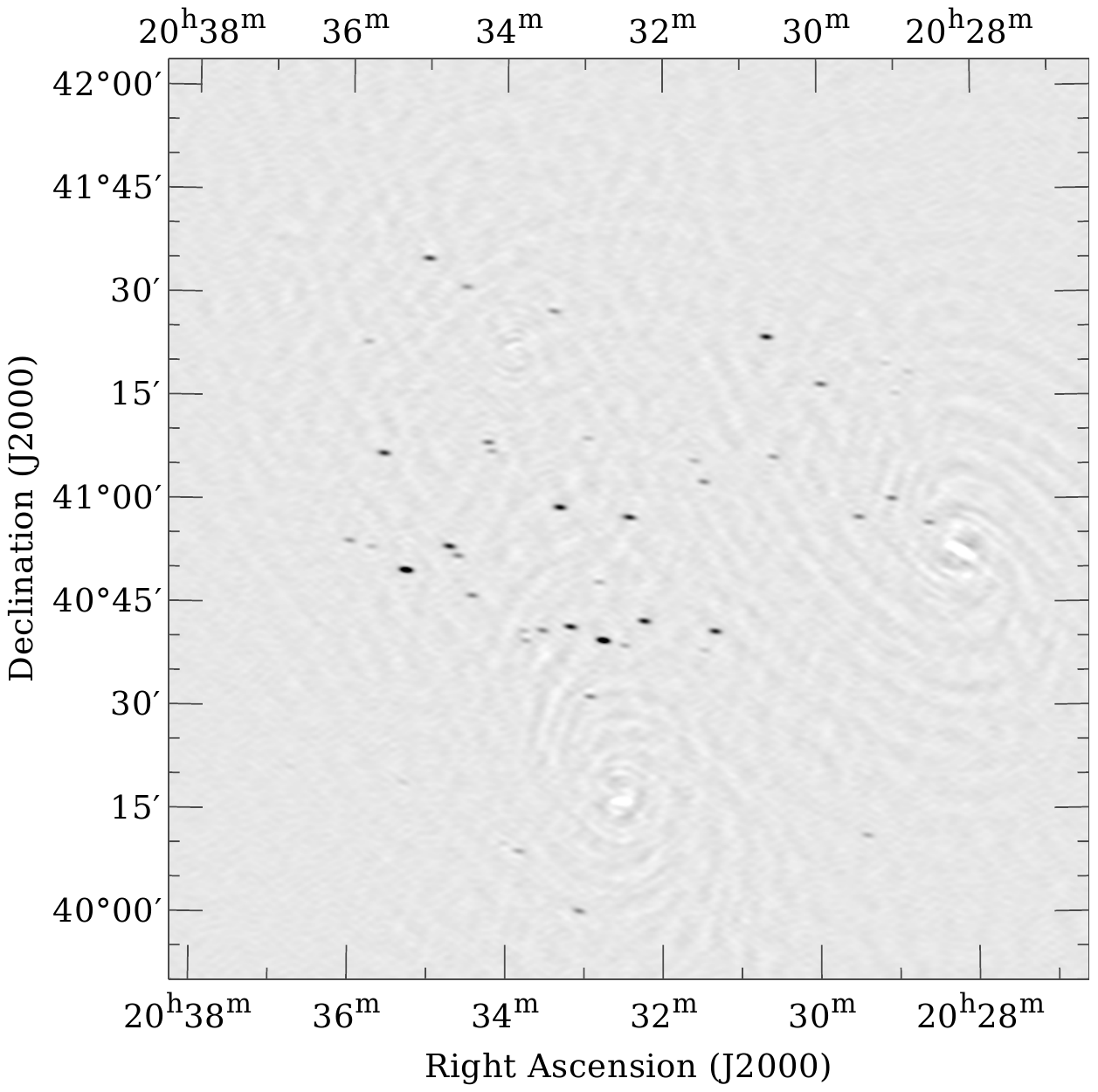}
\caption{Image of the \acyg\ pointing from 2011~Feb~01 with
  LSS subtracted. Primary beam correction has not been applied.
  The grayscale is linear from -5~mJy/beam (white) to
  50~mJy/beam (black). (Note that this is different than that used in
  Figure~\ref{f:deep}.) The maximum brightness in the image is \apx
  380~mJy/beam. The rms deconvolution residual is 0.9~mJy/beam. The
  synthesized beam is $85'' \times 36''$ at a position angle of
  $83\degr$.}
\label{f:singleepoch}
\end{figure}

Figure~\ref{f:singleepoch} shows an image made from a single
\cyg\ epoch (2011~Feb~01) after subtraction of the LSS. The
\cyg\ field contains three bright sources near the half-power point
that are somewhat problematic in the subtraction: \object{DR\,7}
(west), \object{DR\,15} (south), and \object{18P\,61}
\citep[northeast;][]{whl91}. All of these are at least marginally
resolved and embedded in cuspy extended emission. The residuals due
the these sources nonetheless do not significantly impair the imaging
of each epoch. Not shown in the figures, but easily detectable in the
\cyg\ epoch images, are \object{DR\,22} and \object{DR\,21}, as
described above.

We generated a total of 29 epoch images, 23 of the \cyg\ pointing
(dropping six analyzed epochs with dynamic range issues caused by the
2011~March major flare), and 6 of the \kep\ pointing. In
Figure~\ref{f:detlimits} we plot the representative rms of each image
(as reported by \textsf{sfind}) as a function of integration time. In
Figure~\ref{f:epsrcdets}, we show the number of sources detected in
each image.

To assess the effectiveness of the LSS subtraction process, we compare
image rms values in regions with varying levels of LSS flux density. We
selected several source free regions in the \cyg\ and \kep\ fields of
50$\times$100 image pixels. For each epoch image and each region, we
computed the ratio of the rms in that region to the rms in an
equally-sized region on the outskirts of the image, away from all
source emission. Taking this ratio compensates for the varying noise
baseline of each image. Figure~\ref{f:lssrms} summarizes these ratios
as a function of the mean LSS model flux density associated with each
source-free region. (Measurements from the \kep\ field, in which there
is no LSS subtraction, are assigned a mean LSS model flux density of zero.)
The rms ratios increase above the baseline as LSS becomes more
significant, but saturate at a factor of \apx 2 for mean LSS flux densities of
\apx 0.4~mJy. We placed a group of regions especially near the three
bright sources of the \cyg\ field. These display somewhat higher rms
ratios than other regions in areas distant from these bright sources,
but the difference is slight.

To give a sense of the areal coverage of the ASGARD GC field, we show
a preliminary mosaicked image of 3~GHz GC data in
Figure~\ref{f:gpmos}. LSS is not subtracted in this image. The missing
pointings have coverage at lower frequencies but not at 3~GHz. The
Sgr\,A complex, with structured emission reaching brightnesses of
50~Jy/beam, presents a clear challenge. The data contributing to
Figure~\ref{f:gpmos} come from only a few epochs, so the more thorough
hour angle coverage of the complete dataset will make a significant
difference to image quality. Joint deconvolution of some or all of the
pointings might greatly improve the deep map, although this would
depend strongly on how well the ATA primary beam can be modeled.

\subsection{Source Catalog}
\label{s:catalog}

We robustly detect 134 compact sources in the deep images of the two
pointings. There are 86 sources detected in the \kep\ region and 48 in
the \cyg\ region, the difference between the two being primarily due
to the effects of LSS in the latter pointing. The faintest source
discovered in the current \kep\ deep image has a flux density of 2.6~mJy. The
faintest source discovered in the \cyg\ image has a flux density of 9.2~mJy.

Because it is 13.5\degr\ out of the Galactic plane and contains no
LSS, the \kep\ pointing provides a clean testbed for our catalog. All
but one of the sources associated with the \kep\ pointing in our
catalog has a match to a source in the NVSS catalog within 20\arcsec,
and none of these match multiple NVSS catalog entries. The probability
of an individual chance match between the catalogs at this positional
tolerance is \apx 0.5\% \citep{ccg+98}. The unmatched source is
visible, but marginal, in the NVSS imagery. We define a set of NVSS
sources that might be detected in the \kep\ deep image as those that
are projected to have a primary-beam-attenuated (ATA-apparent) flux
density greater than 2.6~mJy, so long as the PB correction factor
$f_\mathrm{PB}$ is smaller than two. Two-thirds (43/66) of these
sources are detected, with most of the nondetections being marginally
discernable in the deep ATA imagery. Meanwhile an additional 41 NVSS
sources are detected that do not meet the above criteria. The
disjunction between the detections is a result of some combination of
spectral dependence, uncertain PB modeling, incompleteness to marginal
detections, and possible genuine source variability. Of the undetected
NVSS sources, the maximum predicted ATA flux density is 5.6~mJy; that
is, there are no NVSS sources that should have easily been detected in
the ATA data that were not, and so there are no bright NVSS sources in
the field that reduced their flux density by a significant fraction by
the time of our observations. All NVSS sources with predicted ATA flux
densities $\ge$5.8~mJy and PB correction factors less than 2 are
detected. If no PB correction limit is applied, the limit is 17.7~mJy,
with a maximum applied PB correction factor of 21.0, corresponding to
an angular separation of 79\arcmin.

NVSS imaging of the \cyg\ region is quite poor and so we do not
compare the catalogs for this region. Instead we use the \sbhw\ L-band
catalog of compact sources in the Cygnus~OB2 region. Of our catalog
entries associated with the \cyg\ pointing, all but 8 are matched to
\sbhw\ sources. Of those, four are outside of the \sbhw\ FOV, one is
clearly detected but is too extended to meet their selection criteria,
and three genuinely do not appear to be detected in the \sbhw\ maps,
in either their L-band or 350~MHz observations. Using the same
criteria as above, one \sbhw\ source might have been expected to be
detected with a predicted apparent ATA flux density of \apx 10~mJy, but it is
blended with the very bright DR\,15. All \sbhw\ sources with predicted
ATA flux densities $\ge$11.2~mJy and PB correction factors less than 2 are
detected.

% 20:31:28.233 +41:02:35.10: pretty clearly in ATA but not SBHW (or
% very faint for them); their fig 8, p. 13 top-right.
% 20:36:00.612 +40:53:53.07: as above, more ambiguous since in some
% LSS; their fig p, p. 14 top-left.
% 20:29:01.114 +41:15:14.75: ditto, fairly convincing; p. 14 topright.
% 20:28:35.644 +40:56:35.01: a little bit extended

In Figure~\ref{f:numbercounts} we compare various source counts. In
order to avoid uncertain primary beam corrections, we restrict our
assessment to sources within the half-power point, reducing our
catalog to 43 and 27 sources within the \kep\ and \cyg\ pointings,
respectively. With such limited numbers, differential functions are
extremely uncertain, and so we report cumulative source count
functions. In the \kep\ pointing, our results are comparable to those
of the NVSS, which is expected given the results reported above. Our
\kep\ counts are somewhat higher than those found in \pgsi, probably
due to a combination of higher-quality deep imaging and manual source
identification, both made possible by the much smaller ASGARD survey
footprint. Our source counts in the \cyg\ pointing are somewhat lower
than those of \sbhw, which may reflect the different observing
frequencies of the two surveys, although such an effect would also be
relevant to the NVSS comparison. Both the ASGARD and \sbhw\ source
counts are uncertain on the 20--50\% level due to small-number
statistics.

\subsection{Detection Limits and Completeness}
\label{s:thresholds}

\begin{figure}[p]
\plotone{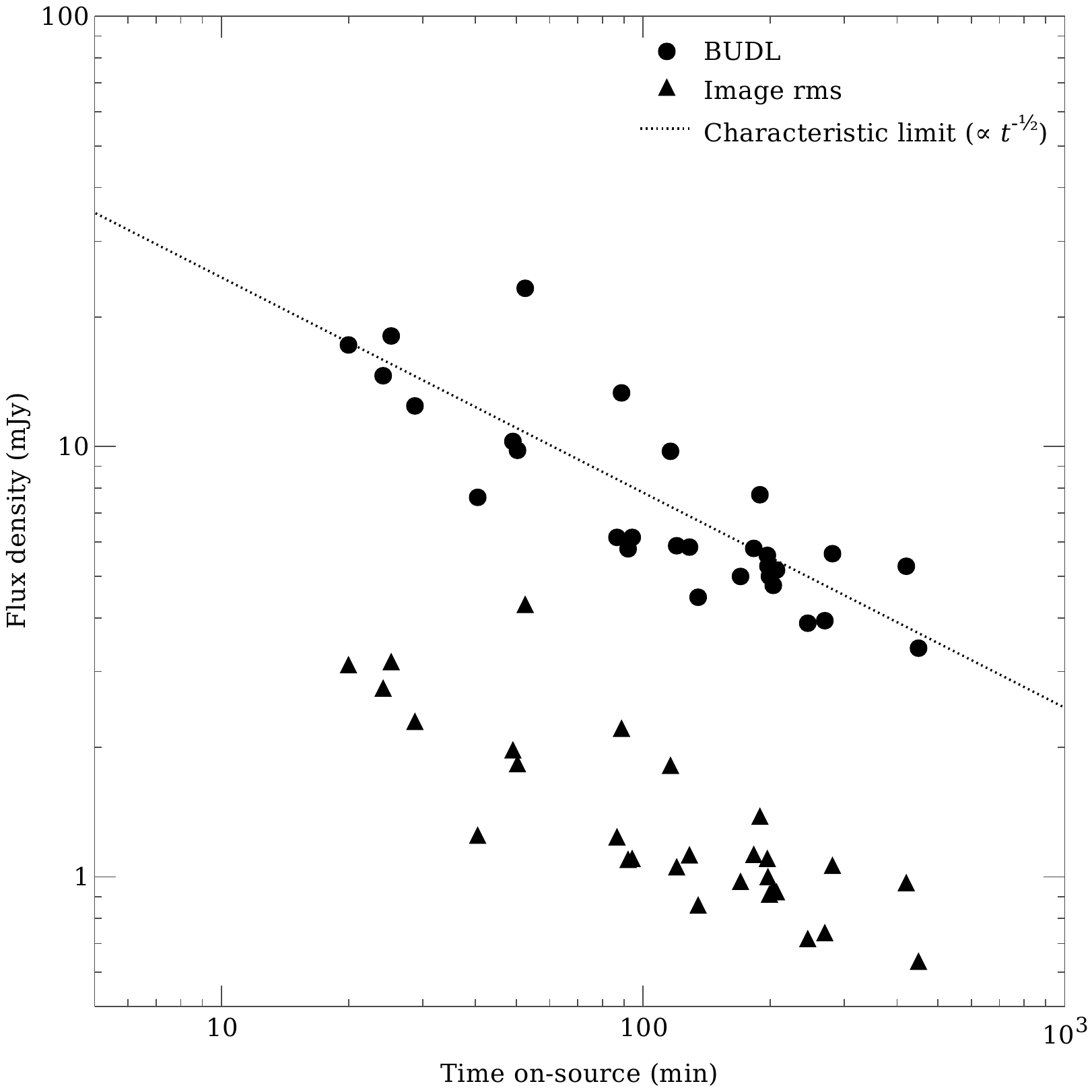}
\caption{Noise statistics for epoch images as a function of on-source
  integration time $t$. \textit{Triangles}: representative rms values
  as reported by \textsf{sfind}. \textit{Circles}: apparent blind
  unresolved-source detection limits (BUDLs; \S\ref{s:thresholds}). At
  fixed $t$, achieved values will vary with system temperature, data
  flagging, calibration quality, and the source detection cutoff
  dynamically determined by the \textsf{sfind} FDR algorithm. The
  \textit{dashed line} shows the characteristic BUDL
  assuming a scaling of $t^{-1/2}$; for $t = 10$~min, this corresponds
  to \apx 24~mJy. Both sets of numbers refer to PB-attenuated images
  and hence \textit{apparent} flux densities.}
\label{f:detlimits}
\end{figure}

\begin{figure}[p]
\plotone{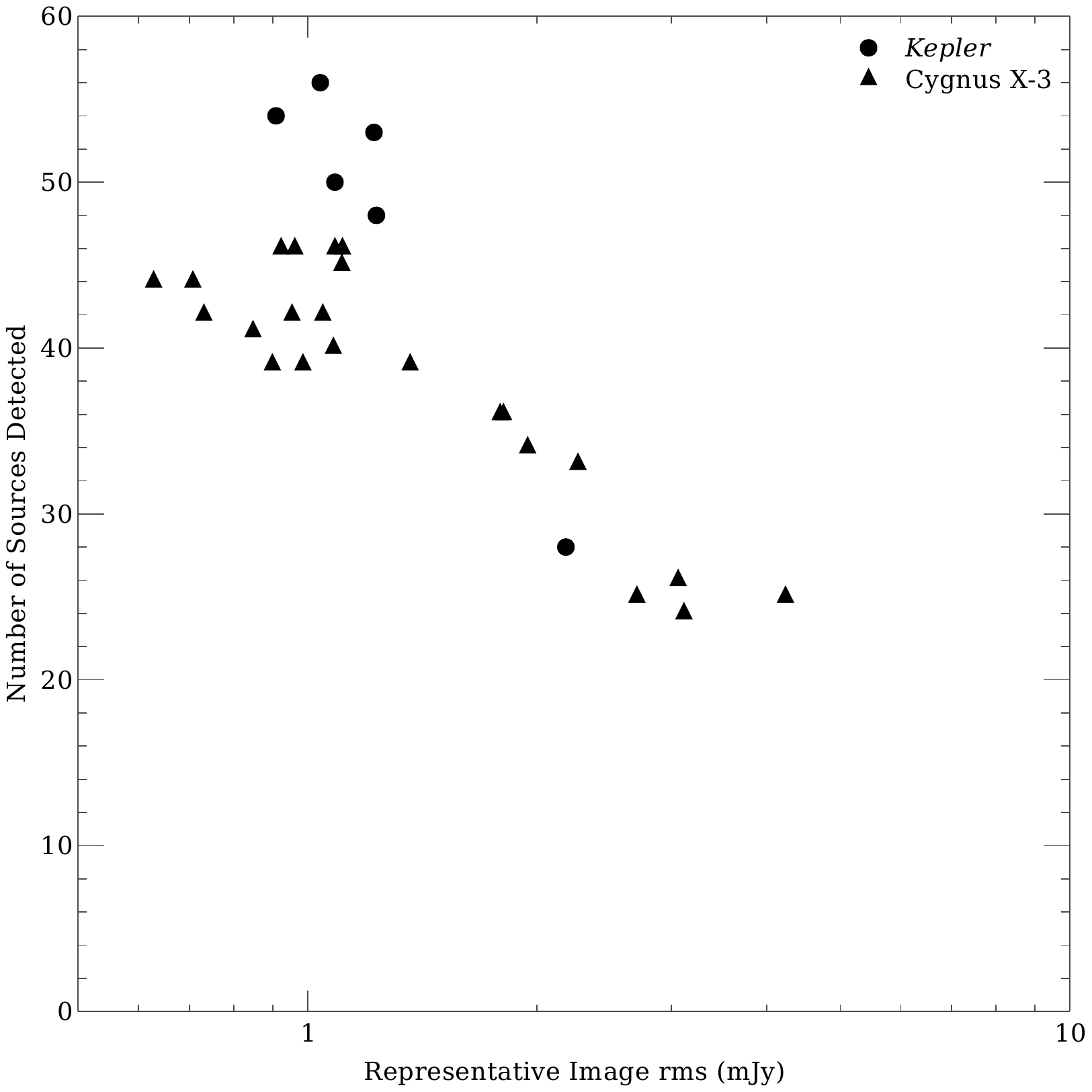}
\caption{Number of sources detected in each epoch image, as a function
  of the representative image rms reported by \textsf{sfind}. Symbols
  are grouped by pointing. There are 86 cataloged sources associated
  with the \kep\ pointing and 48 with the \acyg\ pointing. The
  detection rate for the \kep\ pointing is lower because the
  \kep\ deep image is much less limited by systematics than the
  \acyg\ deep image, and so contains many more faint sources that are
  cataloged but cannot easily be detected in the epoch images.}
\label{f:epsrcdets}
\end{figure}

\begin{figure}[p]
\plotone{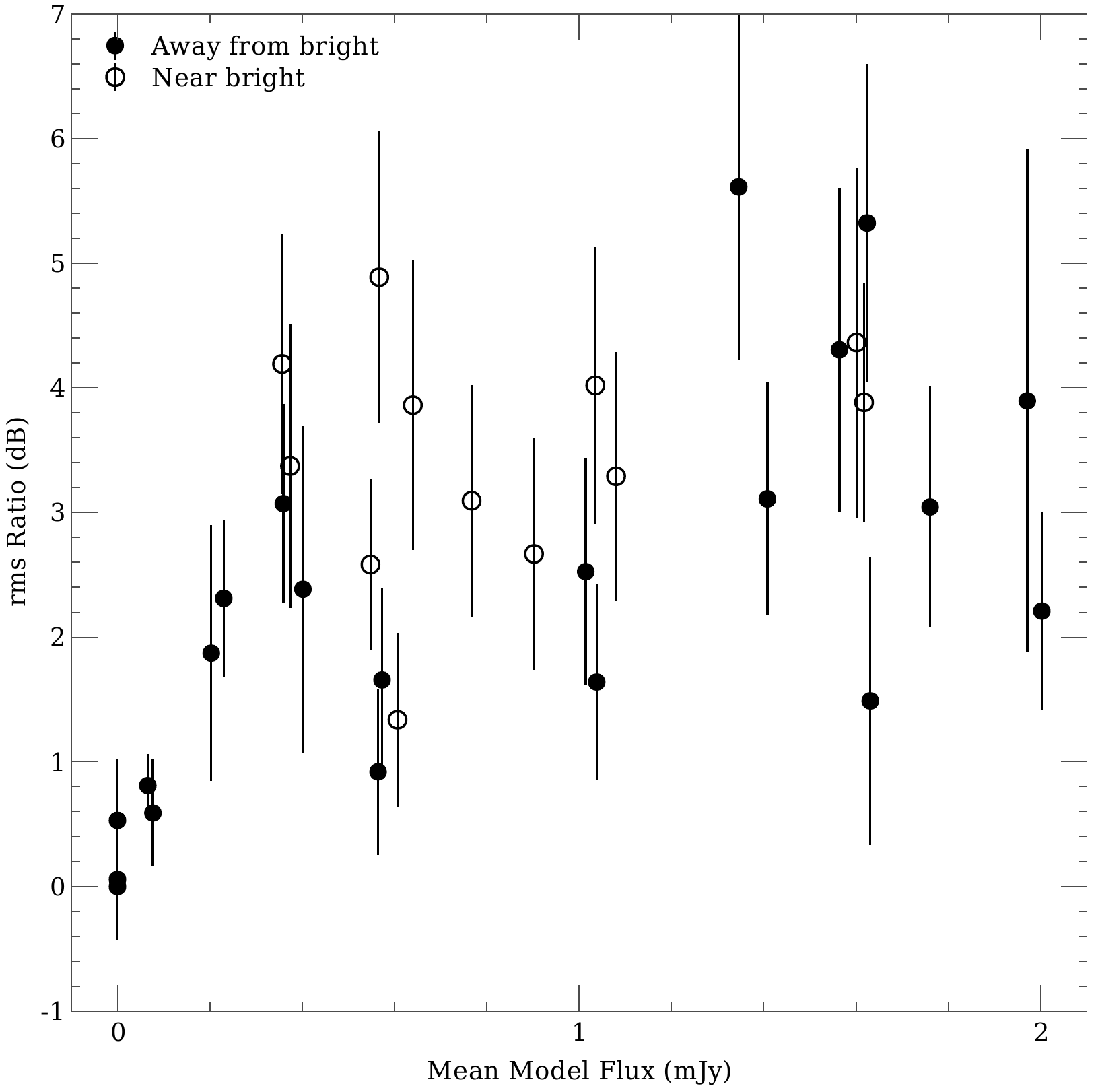}
\caption{Measurements of the increase in image noise as a function of
  LSS flux density. Source-free regions of 50$\times$100 pixels were selected
  in the \kep\ and \acyg\ fields. For each epoch image and each
  region, the rms was computed and divided by the rms in an
  equally-sized box in the lower left corner of each image, far away
  from all source emission. Because these values are ratios, they are
  independent of the primary beam correction.
  Each plotted point gives the mean and
  standard deviation of these rms ratios over all imaged epochs for a
  particular pointing, source-free region, and focus setting. The
  \textit{abscissa} is the mean flux density of the LSS model of that
  region. \textit{Open circles} denote measurements from regions very
  near (\apx 20 pixels) any of the three bright sources of the
  \acyg\ field. Areas in which there is significant LSS show rms
  increases of a factor of \apx 2 versus the source-free baseline. See
  \S\ref{s:images} for discussion.}
\label{f:lssrms}
\end{figure}

\begin{figure}[p]
\plotone{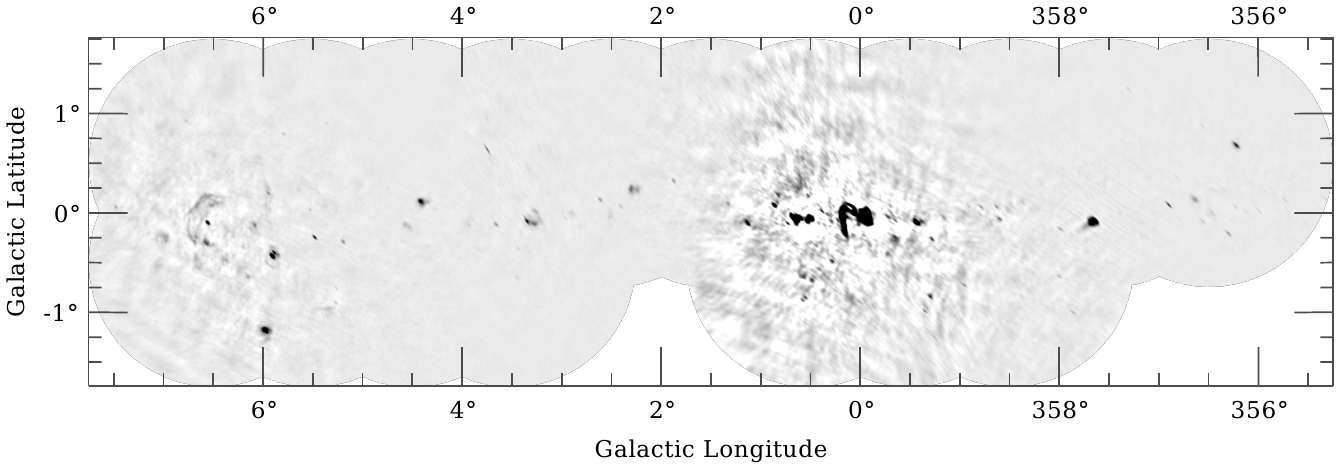}
\caption{Preliminary mosaicked image of ASGARD GC field. Unlike the
  other maps in this paper, a primary beam correction has been
  applied. The
  grayscale is linear from -70~mJy/beam (white) to 800~mJy/beam
  (black). The maximum brightness in the image is \apx
  3920~mJy/beam. Each pointing has been imaged to a diameter of about
  2.2 times the HPBW. Most pointings have only a single epoch of
  observations contributing to the image.}
\label{f:gpmos}
\end{figure}

\begin{figure}[p]
\plotone{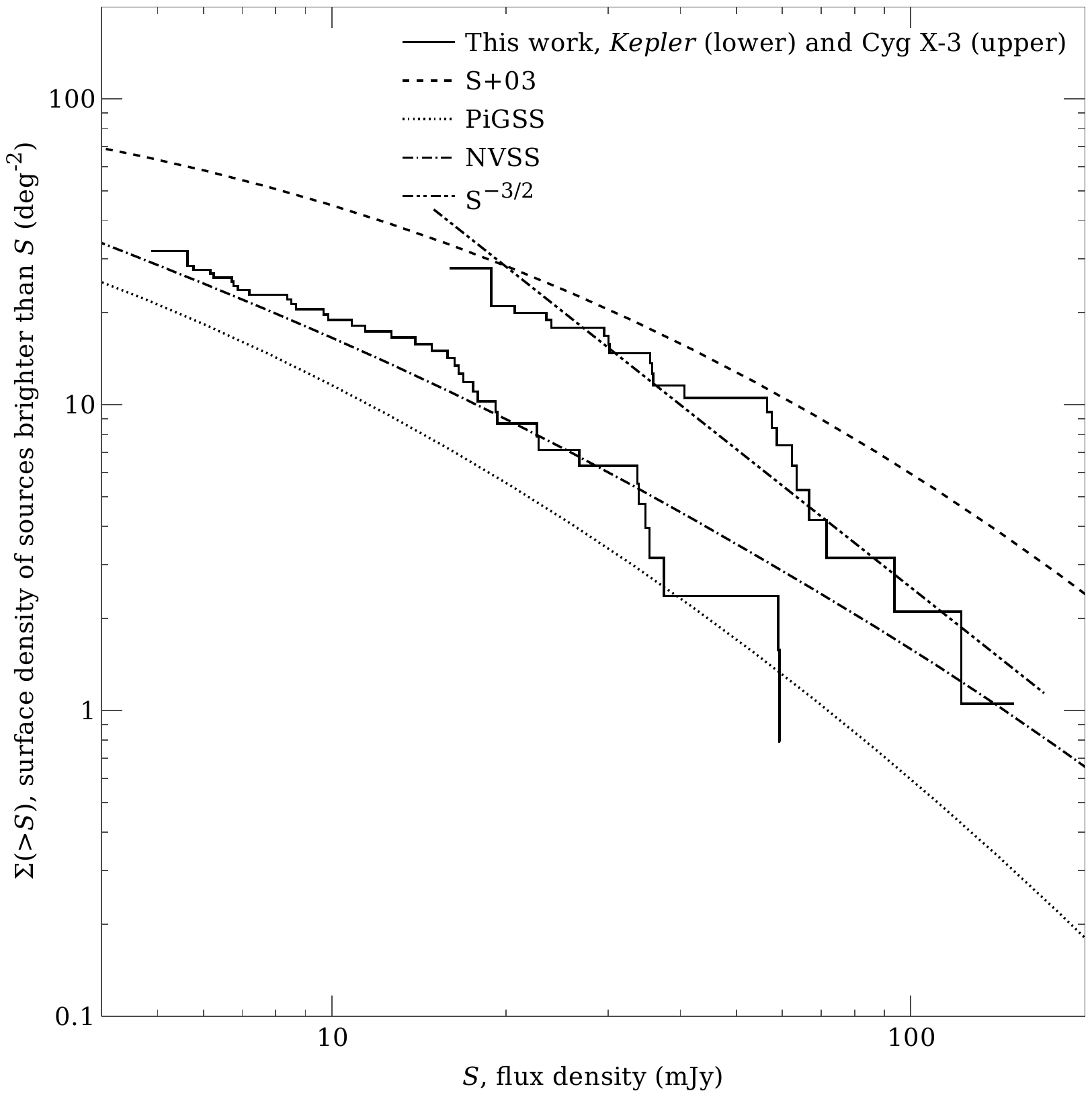}
\caption{Cumulative source counts in this and other surveys. ASGARD
  measurements are derived from the cumulative distribution function
  of source flux densities as detected in the \textit{Kepler} and
  \acyg\ deep images, combined with effective search areas based on
  our analytic primary beam models. Reference values are derived from
  differential source count measurements reported in
  \citetalias{sgdbvdh+03}, the NVSS \citep{ccg+98}, and \pgsi. The
  arbitrarily-normalized $S^{-3/2}$ line shows the expected scaling
  for a Euclidean, volume-limited distribution. See \S\ref{s:catalog}
  for discussion.}
\label{f:numbercounts}
\end{figure}

In order to analyze searches for transient and highly-variable
sources, we must understand the completeness of these searches. The
relevant detection limit for transient searches such as our
\textsf{sfind} step (\S\ref{s:sfind}) is, by definition, the
\textit{blind} source detection limit. Furthermore, because the ATA
will not resolve true astrophysical transients, the detection limit
may be expressed as a flux density limit rather than a brightness
limit. As a shorthand we thus refer to this particular quantity as the
``BUDL'': blind unresolved-source detection limit. If artifacts are
not significant, an image that has not been corrected for primary beam
attenuation has a single value of the BUDL in terms of
\textit{apparent} flux density: the limit reported by
\textsf{sfind}. Figure~\ref{f:detlimits} depicts the variation of this
value as a function of integration time for the epoch images analyzed
in this work. The BUDL in terms of \textit{intrinsic} flux density
varies spatially as defined by the primary beam response. The apparent
BUDL is a function of each image's thermal noise; the additional
effective noise due to limitations in calibration, deconvolution,
etc.; and the source detection cutoff determined by the \textsf{sfind}
FDR algorithm. Typically, the apparent $\mathrm{BUDL} \apx 5.1\sigma$,
where $\sigma$ is the representive image rms, but the coefficient
varies in an image-dependent manner due to the use of the FDR
algorithm. As shown in Figure~\ref{f:detlimits}, however, its range of
variation is not very large. We generally find that $\mathrm{BUDL}
\propto \tau^{-1/2}$ as would be expected, where $\tau$ is the image
integration time.

We insert artifical sources into our imagery and then apply our
source-finding method in order to find an upper limit to our method's
completeness as a function of flux density.
This approach only finds an upper limit to
the completeness because we insert the sources mid-way through our
processing pipeline, after the \uv\ data have been initially
calibrated. We experimented with inserting false sources in both the
\uv\ and image domains and obtained equivalent results with both
approaches, so for more detailed studies we used image-domain
insertion, which avoids the need to rerun the
computationally-expensive imaging steps.

\begin{figure}[p]
\plotone{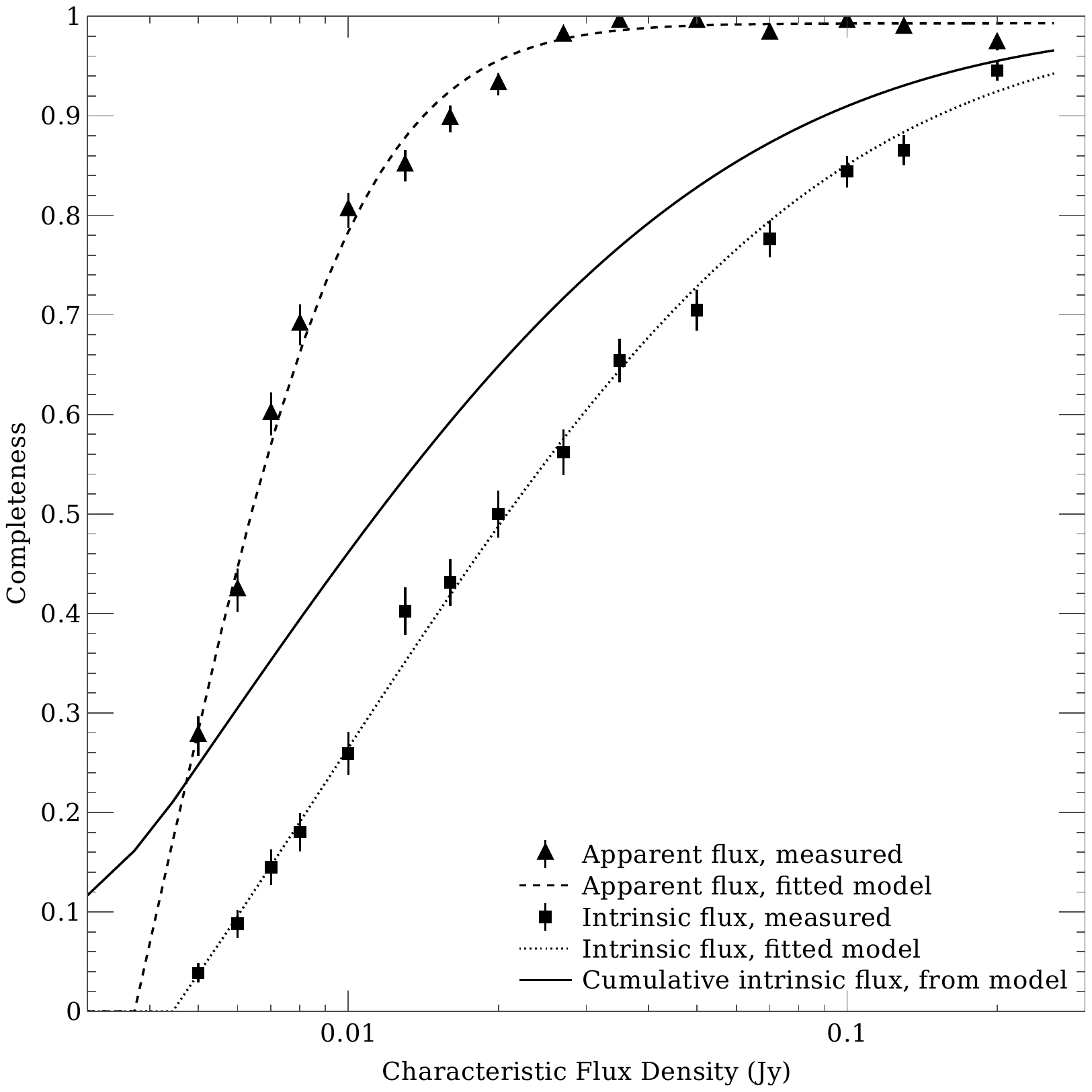}
\caption{Completeness of blind source detection in various regimes, as
  determined from simulations with the current set of epoch images.
  \textit{Apparent} completeness measures the detection fraction of
  sources of a given flux density in images without correction for primary
  beam attenuation. \textit{Intrinsic} completeness measures the
  detection fraction of sources of a single given flux density after
  correction for PB attenuation. \textit{Cumulative intrinsic}
  measures the expected detection fraction for all sources brighter
  than a given intrinsic flux density, assuming luminosity function $N({>}S)
  \propto S^{-3/2}$. See \S\ref{s:thresholds} for discussion.}
\label{f:completeness}
\end{figure}

We began by sampling the differential completeness at fixed
\textit{apparent} flux density $c_a(S_a)$, that is, without accounting for
primary beam attenuation. To sample this function, we spread \apx 500
unresolved sources of flux density $S_a$ throughout all of the epoch
images, spacing them randomly but evenly in position angle and radius
relative to the pointing center to avoid overlaps. (This is fair
because we expect any image to contain at most one transient.) The
blind source detection routines are then run and the fraction of
detected sources is reported as the completeness, assigning an
uncertainty to the measurement assuming Poisson statistics. The top
curve of Figure~\ref{f:completeness} shows the results of sampling $4
< S_a < 200$~mJy and a four-parameter analytic fit to these
results. Our model function is
\begin{align}
\tilde c_a(S_a) = \begin{cases}
0 & \text{if $S_a < (-D / B)^{1/C}$}, \\
A \tanh(B S_a^C + D) & \text{otherwise}.
\end{cases}
\end{align}
There is no particular theoretical motivation for this representation,
but it can be made to match the data well empirically and has two
desirable analytical traits. Firstly, it plateaus as $S_a
\to \infty$. Secondly, it has realistic discontinuous behavior at the
completeness zero point. The fitted parameter $A$, the plateau value,
is 99.3\%, and reflects that a small fraction of our images is
disqualified from blind source detection by virtue of proximity to a
cataloged source.

We also sample the differential \textit{intrinsic} completeness
function $c_i(S_i)$, assuming our analytic PB model, by inserting
false sources with appropriately PB-attenuated flux densities. Our results are
also shown in Figure~\ref{f:completeness}, along with a corresponding
fitted model of the completeness function in which the plateau value
was fixed to match that of the apparent completeness function. There
is substantial disagreement between our samples of this function and
an estimate derived from $c_a$ and our image BUDLs, suggesting that
the assumption of a spatially-uniform BUDL in each image does not hold
strongly.

Finally, we compute the \textit{cumulative} intrinsic completeness
$C_i$; that is, the completeness to all sources intrinsically brighter
than some limiting value. In order to do this we must assume a
distribution function for source number counts as a function of
flux density. We use the standard Euclidean, volume-limited distribution
$N({>}S) \propto S^{-3/2}$, which agrees well with the observed number
counts of our catalog (Figure~\ref{f:numbercounts}). The cumulative
intrinsic completeness is then:
\begin{align}
C_i({>}S_i) &= \int_{S_i}^\infty \mathrm{d}S_i' c_i(S_i') (S_i')^{-5/2}
 \left/ \int_{S_i}^\infty \mathrm{d}S_i' (S_i')^{-5/2}\right. \\
 &= \frac{3}{2} S_i^{3/2} \int_{S_i}^\infty \mathrm{d}S_i' c_i(S_i') (S_i')^{-5/2}.
\end{align}
We evaluate this integral numerically using the analytic fit to our
samples of $c_i(S_i)$. Below the cutoff of $c_i$ at 4.4~mJy,
$C_i({>}S_i) \propto S_i^{3/2}$.

\subsection{Search for Transients}
\label{s:transients}

We performed a search for transient radio sources with our subsample
of processed ASGARD images. Recall that by our definition,
``transient'' sources are merely ones that are not detected in our
deep images, and so must be searched for in the individual epoch
images. Sources detected in this way will, by construction, evolve on
relatively short timescales, but it is important to note that we are
also sensitive to sources that vary on relatively long timescales:
it is just that they are detected in the deep images, so they do not
satisfy our definition of transience. Once detected, any
transient sources are entered into our catalog and their photometry is
measured at every epoch, so both classes of sources are analyzed
identically, although in the stereotypical single-epoch case the
``lightcurve'' for a transient will consist of a numerous upper limits
and a single detection.

After applying the \textsf{sfind} source detection and filtering
procedure described in \S\ref{s:sfind}, we were left with a list of 17
transient candidates in our 29 epoch images (23 of the \cyg~pointing
excluding of the 2011~March major flare, six of the
\kep\ pointing). All but one were detected in the
\cyg\ pointing. Visual inspection confirmed that all of the candidates
were spurious, as indicated by various combinations of unphysical
extended structure, a large distance from the pointing center,
association with sidelobes or incompletely subtracted LSS, and/or
indistinguishability from other noise fluctuations in the image. The
full-scale ASGARD transient search will verify transient candidates
rigorously, both by partitioning the imaged data to check for
instrumental errors \citep[cf.]{fko+12} and by treating the image
noise statistics quantitatively.

Although experience suggests that systematic effects are far more
likely to cause false detections than thermal fluctuations, it is
instructive to consider the limits imposed by noise
\citep{fko+12}. Our search examined a total area of \apx$7\times 10^5$
synthesized beams, accounting for the variable beam size and HPBW of
each image. This corresponds to one expected noise event of
4.68$\sigma$ assuming purely Gaussian statistics. All of our
candidates are above this threshold, so our search is not yet
contaminated by statistical fluctuations. Extrapolating to the
complete 3~GHz dataset, one statistical noise fluctuation of \apx
5.4$\sigma$ may be expected.

We quantify the power of this early search by computing its effective
search area as a function of the BUDL. Given the BUDL of an image
$S_{a,\mathrm{lim}}$, our analytic primary beam model, and the
cumulative apparent-flux-density completeness function $C_i(S_i)$ determined
in the previous section, the completeness-corrected effective solid
angle in which sources with intrinsic flux density greater
$S_{i,\mathrm{lim}}$ may be detected is
\begin{equation}\Omega_\mathrm{eff}(S_{a,\mathrm{lim}},{>}S_{i,\mathrm{lim}}) =
C_i({>}S_{i,\mathrm{lim}}) \pi
r^2(S_{a,\mathrm{lim}},S_{i,\mathrm{lim}}),
\end{equation}
where $r$ is the radius
within which every source of intrinsic flux density ${>}S_i$ will be
detected, given a fixed BUDL:
\begin{align}
r^2(S_a,S_i) = \begin{cases}
0 & \text{if $S_i < S_a$}, \\
2\sigma_\mathrm{PB}^2 \log (\frac{S_i}{S_a}) & \text{if $S_i < 90 S_a$}, \\
(3\sigma_\mathrm{PB})^2 & \text{otherwise}.
\end{cases}
\end{align}
(The upper limit stems from our rejection of \textsf{sfind} sources
found beyond the 98.9\% attenuation point.) Following \citet{bsb+07},
a transient survey consisting of $N$ visits to the same field of solid
angle $\Omega$ probes a total area of \apx $\Omega(N-1)$, if the time
elapsed between visits is larger than the transient timescale. If the
noise in each epoch varies, the effective area searched in each epoch
also varies, given a fixed $S_{i,\mathrm{lim}}$. Denoting these areas
$\Omega_i$, where $\Omega_1 > \Omega_2 > \ldots > \Omega_N$, the
effective area searched in this case is $\sum_{i=2}^N\Omega_i$, if the
area probed by each epoch is a strict subset of the area probed by
every more sensitive epoch.

The above condition holds in the case of identical pointing centers
and nonincreasing radial primary beam shapes, but not for mosiacs with
overlapping pointings. The mosaic case can be treated numerically by
explicitly mapping the areal contribution of each epoch. We take this
approach to determine the effective area as a function of
$S_{i,\mathrm{lim}}$ for our processed data. Using the same approach
as \citet{fko+12}, these measurements can be converted into upper
limits on snapshot transient source areal densities. If a total area
of $\Omega_\mathrm{eff}$ is searched and no transients are detected,
the 95\% confidence limit (CL) density upper limit is $3 /
\Omega_\mathrm{eff}$ to a very good approximation. Our results are
plotted in Figure~\ref{f:sadlimits} and discussed in \S\ref{s:conc}.

We also extrapolate this technique to anticipate the parameter space
that will be probed by the full survey. We use the empirical
relationship between integration time and BUDL shown in
Figure~\ref{f:detlimits} to estimate the limit for all epochs as yet
unprocessed. When doing so we add in a scatter comparable to that seen
in the figure. This approach ``bakes in'' typical levels of data loss
due to RFI, instrumental malfunctions, and so on, assuming that the
data processed thus far are not unusual in those regards. Compared to
what is presented in this work, the full dataset will be more powerful
at the bright/rare end because most of the longest integrations have
already been processed.

\subsection{Variability Analysis}
\label{s:var}

Several different metrics are commonly used for assessing the
variability of radio sources. Most techniques compute a single scalar
variability metric for each source, although this approach is
necessarily reductive. (For instance, a source may be highly variable
on one timescale and less so on another.) Analyses such as the
structure function approach \citep{sch85,emhu10} are more
sophisticated, although they still encode certain assumptions about
the nature of the variability being measured. Well-designed scalar
metrics can still, however, capture meaningful information about
overall variability, as we describe below.

\citet{bhw+10} define a modulation index or ``fractional variation''
\begin{equation}
f = \frac{S_\mathrm{max}}{S_\mathrm{min}},
\label{e:modindex}
\end{equation}
but as discussed by \citet{ofb+11} this metric has irregular
statistical properties depending on the number of epochs of
observations. The same is true of the metric
\begin{equation}
V = \frac{S_\mathrm{max} - S_\mathrm{min}}{S_\mathrm{max} +
S_\mathrm{min}}
\end{equation}
used by \citet{gt86}, which is algebraically interchangeable with the
above. \citet{ofb+11} prefer the ratio of the standard deviation of
the flux density measurements to the weighted mean, written in their notation
as ``$\mathrm{StD}/\langle f \rangle$'' (and sometimes also referred
to as ``the'' modulation index). In our notation this would be
$\sigma_S / \smn$, where the maximum-likelihood mean flux density \smn\ of a
set of $n$ measurements $S_i$ with associated uncertainties $\sigma_i$
is
\begin{equation}
\smn = \sum_{i=1}^n S_i \sigma_i^{-2} \left/ \sum_i \sigma_i^{-2} \right..
\end{equation}
While $\sigma_S / \smn$ is useful for comparing different studies, we
disfavor it for the ranking of candidate variables because it is
insensitive to the scale of the uncertainties in each measurement, and
thus the confidence with which varying and nonvarying sources can be
distinguished. (This is easy to see if the ratio is expressed as
\begin{equation}
\frac{\sigma_S}{\smn} = \frac{\sqrt{\langle S^2 \rangle - \smn^2}}{\smn},
\end{equation}
where the uncertainties only appear implicitly in the averages.)

Building on \pgsii\ and \citet{bmg+11errated} we prefer $\chi^2$
statistics for this purpose. The $\chi^2$ statistic regarding the
hypothesis that some source is unvarying is
\begin{equation}
\chi^2 = \sum_{i=1}^n\left(\frac{S_i - \smn}{\sigma_i}\right)^2.
\end{equation}
The distribution of observed $\chi^2$ values among the observed
sources is not well-defined because, as alluded to in
\citet{bmg+11errated}, they are not drawn from one parent
distribution. They rather come from the family of $\chi^2_k$
distributions, where $k = n - 1$ is the number of degrees of freedom
associated with each source. Computing a reduced $\chi^2$ does not
help because that only normalizes the expectation values of the
distributions, not their shapes. To obtain a well-defined distribution
we must instead apply the full cumulative distribution function (CDF)
of the $\chi_k^2$ family, computing for each source $P_c$, the
probability of accepting the hypothesis that it is constant. We give
the expression for $P_c$ in Appendix~\ref{a:pc}
(Equation~\ref{e:pc}). In the case of Gaussian errors and no
variability, the observed $P_c$ values of all sources will be
uniformly distributed between 0 and 1.

The metrics listed above do not use the timestamps associated with
each flux density measurement. Therefore, although they can provide an
overall assessment of whether a source is somehow ``variable,'' they
cannot describe the nature of that variability. Among this set of
metrics $P_c$ is in some sense ideal because it is precisely a
probabalistic assessment of this matter.  \citet{ofb+11} find that the
structure function of Galactic radio variables saturates at $\tau \apx
10$~d, remaining flat to at least $\tau \apx 60$~d, suggesting that
our basic rankings are useful. Nonetheless for particular sources
structure functions can, for instance, probe the contribution of
scintillation to the observed variability \citeeg{rcm00,lrm+08}. The
many epochs of ASGARD provide a rich dataset for this analysis: the
3~GHz dataset spans a total time baseline of 1.2~yr and contains 18
pointings with more than twenty visits and 14 pointings with more than
fifty.

Sources farther away from the pointing center tend to have increasing
systematic effects. This can be seen by examining trends in $P_c$ or
$\sigma_S/\smn$ against the PB correction factor $f_\mathrm{PB}$, as
seen in Figure~\ref{f:varvspbfactor}: although genuine variability
will not increase with distance from the pointing center, apparent
variability does. In Figure~\ref{f:pcbypbf}, we compare the $P_c$ CDFs
of sources inside and outside of the half-power radius. In the latter
population, there is an overabundance of both apparent variability and
sources with overestimated uncertainties
(cf. Appendix~\ref{a:pc}). This leads us to restrict our variability
analysis to sources within the half-power point, which constitute 53
of the 97 sources that have sufficient detections to assess
variability.

The excess on the low end of the $P_c$ CDF may be indicative of the
presence of genuinely variable sources, but the sample size is
insufficient to allow conclusive determination. For the same reason we
do not attempt to evaluate a variability confidence threshold as
computed by \citet{bmg+11errated}. More generally, the irregularity in
the CDF and its divergence from the theoretically-motivated shape
defined by Equations~\ref{e:cdffirst}--\ref{e:cdflast} indicate that
our $P_c$ values are not statistically rigorous, although they are
useful for ranking the variability of sources. For instance, in
Figure~\ref{f:varvsflux} we compare the dependence of $P_c$ and
$\sigma_S / \smn$ on PB-corrected source flux density. For the more reliable
sources with $f_\mathrm{PB} < 2$, there is a clear increase in
$\sigma_S / \smn$ as \smn\ decreases. As mentioned above, we attribute
this to the fact that $\sigma_S / \smn$ does not account for the
overall scale of the uncertainties in a set of measurements, which are
fractionally larger for fainter sources. The $P_c$ metric does not
show this flux density dependence.

In Table~\ref{t:vars} we present positions and variability metrics of
the eight most variable sources, as ranked by $P_c$, in the subset of
sources found within the half-power point. Lightcurves for these
sources are plotted in Figure~\ref{f:varlc}, and image cutouts showing
the sources as imaged are shown in Figures~\ref{f:stampscyg} and
\ref{f:stampskep}. We discuss the nature of these sources in
\S\ref{s:varsrcs}. \cyg\ is the most variable source and is
emphatically detected as so by every metric, even without the
inclusion of the 20~Jy measurements from its 2011~March major
flare. None of the other sources ranked as highly-variable are
obviously so. Six of the eight highly-variable sources are found in
the \kep\ pointing, and four of these have nearby companions (see
\S\ref{s:varsrcs}). Although our photometry routines simultaneously
fit nearby sources, it is possible that undeconvolved sidelobes are
affecting our flux density measurements. Epochs with low-quality images tend
to be more obviously subject to systematic photometric effects. In our
investigations, post-imaging calibration of these images has been
unable to remove these effects: although certain sources are
systematically shifted to lower flux densities, different sources are
systematically shifted upward, so a simple scale factor is an
insufficient correction.

\section{Discussion \& Conclusions}
\label{s:conc}

Our first results demonstrate the characteristics of the ASGARD
dataset, the strategies we have used in our processing pipeline, and
an initial search for highly variable and transient sources.

\subsection{Galactic Radio Transient Areal Densities}
\label{s:sadlimits}

\begin{figure}[p]
\plotone{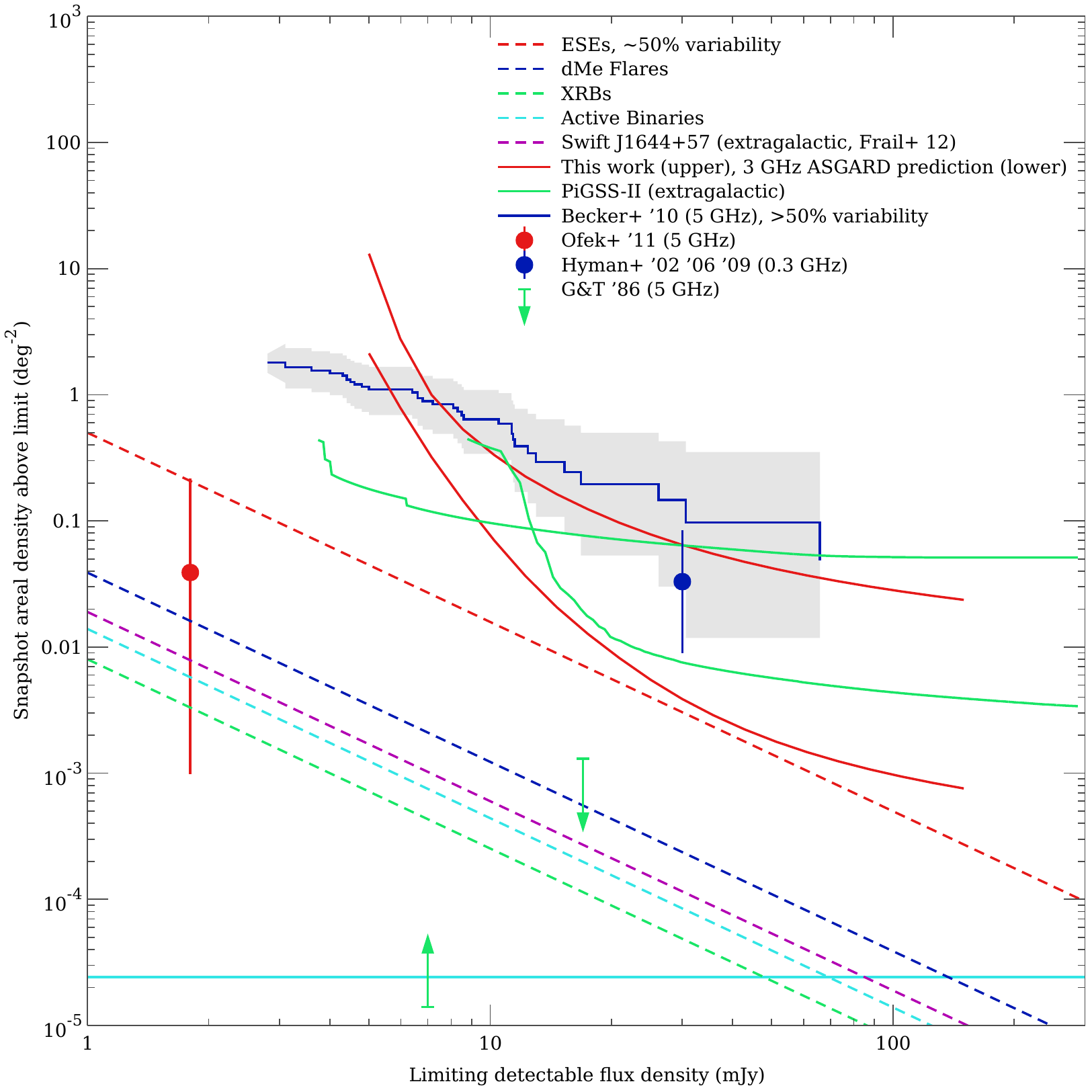}
\caption{Measurements of and limits to snapshot areal densities of
  radio transients. See \S\ref{s:sadlimits} for detailed
  discussion. All uncertainties are to 95\% confidence
  limit. \textit{Upper red line}: upper limit on the areal density of
  Galactic transients from this work alone. \textit{Lower red line}:
  predicted upper limit from the complete 3~GHz ASGARD dataset,
  should no Galactic transients be found. \textit{Blue line}: areal
  density of Galactic sources with fractional variability > 50\% at
  5~GHz on multi-year timescales as measured by
  \citet{bhw+10}. \textit{Green lines}: extragalactic transient areal
  density limits from \pgsii\ on daily and monthly timescales. See
  \citet{bwb+11} for more information.
  \textit{Points and limit arrows}: measurements of
  Galactic transient rates from the VLA at 5~GHz \citep{ofb+11}, VLA
  at 0.33~GHz \citep{hlk+02,hlr+06}, GMRT
  at 0.235~GHz \citep{hwl+09}, and NRAO 91-m transit telescope at
  5~GHz \citep{gt86}, as discussed in the text. \textit{Dashed lines}:
  areal density estimates for various radio transient populations as
  discussed in the text. \textit{Horizontal line}: density
  corresponding to one event on the whole sky at any given time.}
\label{f:sadlimits}
\end{figure}

\begin{figure}[p]
\plotone{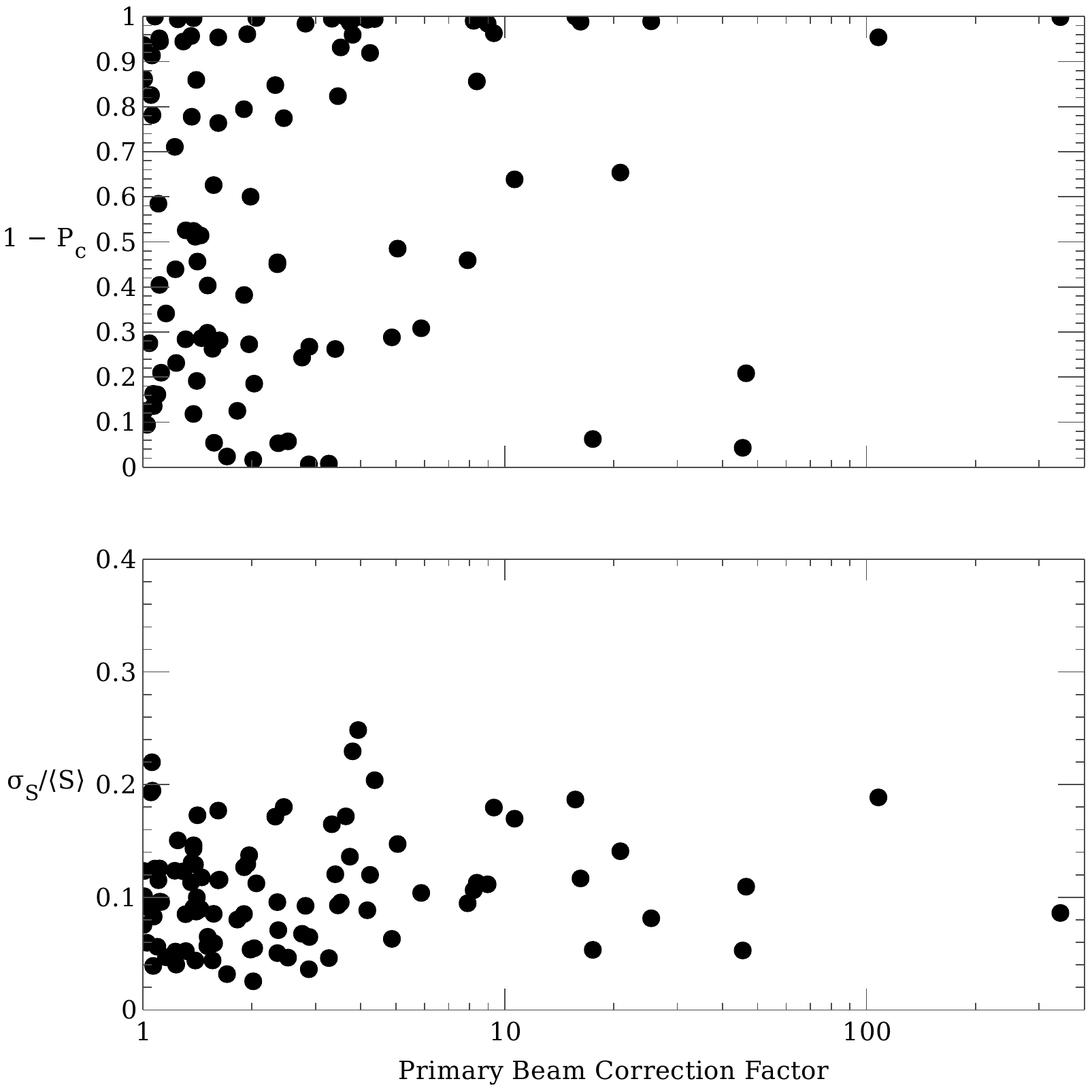}
\caption{Variability metrics as a function of primary beam correction
  factor $f_\mathrm{PB}$. The upper panel plots $1 - P_c$ so that
  a higher vertical position indicates more variability in both
  panels. The distributions of the metrics both skew toward higher
  values as $f_\mathrm{PB}$ increases, suggesting less-reliable
  measurements. Cyg~X-3 is omitted from these plots. It has $1 - P_c
  \approx 1$ and $\sigma/\smn \approx 0.69$.}
\label{f:varvspbfactor}
\end{figure}

\begin{figure}[p]
\plotone{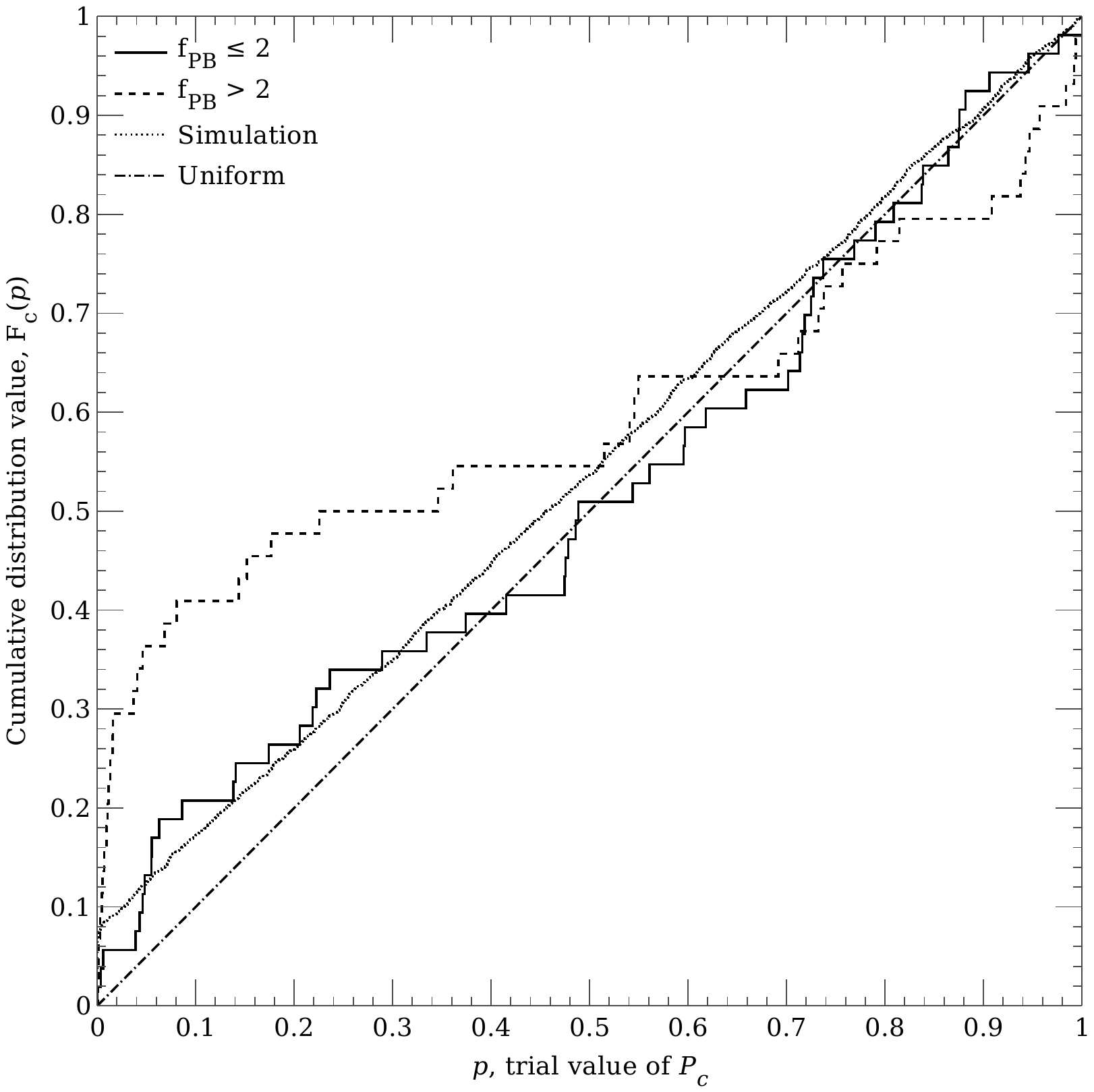}
\caption{Cumulative distribution of observed $P_c$ values for sources
  inside and outside the half-power point: given a randomly-chosen
  source in the dataset, $F_c(p)$ is the probability that its $P_c$
  value is less than $p$. \textit{Dash-dotted line}: the uniform
  distribution, $F_c(p) = p$, which would be expected in the absence
  of variability with purely Gaussian errors. \textit{Dotted line}:
  simulated observations based on our sampling and a population in
  which 10\% of sources have log-normal flux density variability with a
  scatter of 0.13~dex. \textit{Solid (dashed) line:} distribution of
  values inside (outside) the half-power point. See \S\ref{s:var} and
  Appendix~\ref{a:pc} for discussion.}
\label{f:pcbypbf}
\end{figure}

\begin{figure}[p]
\plotone{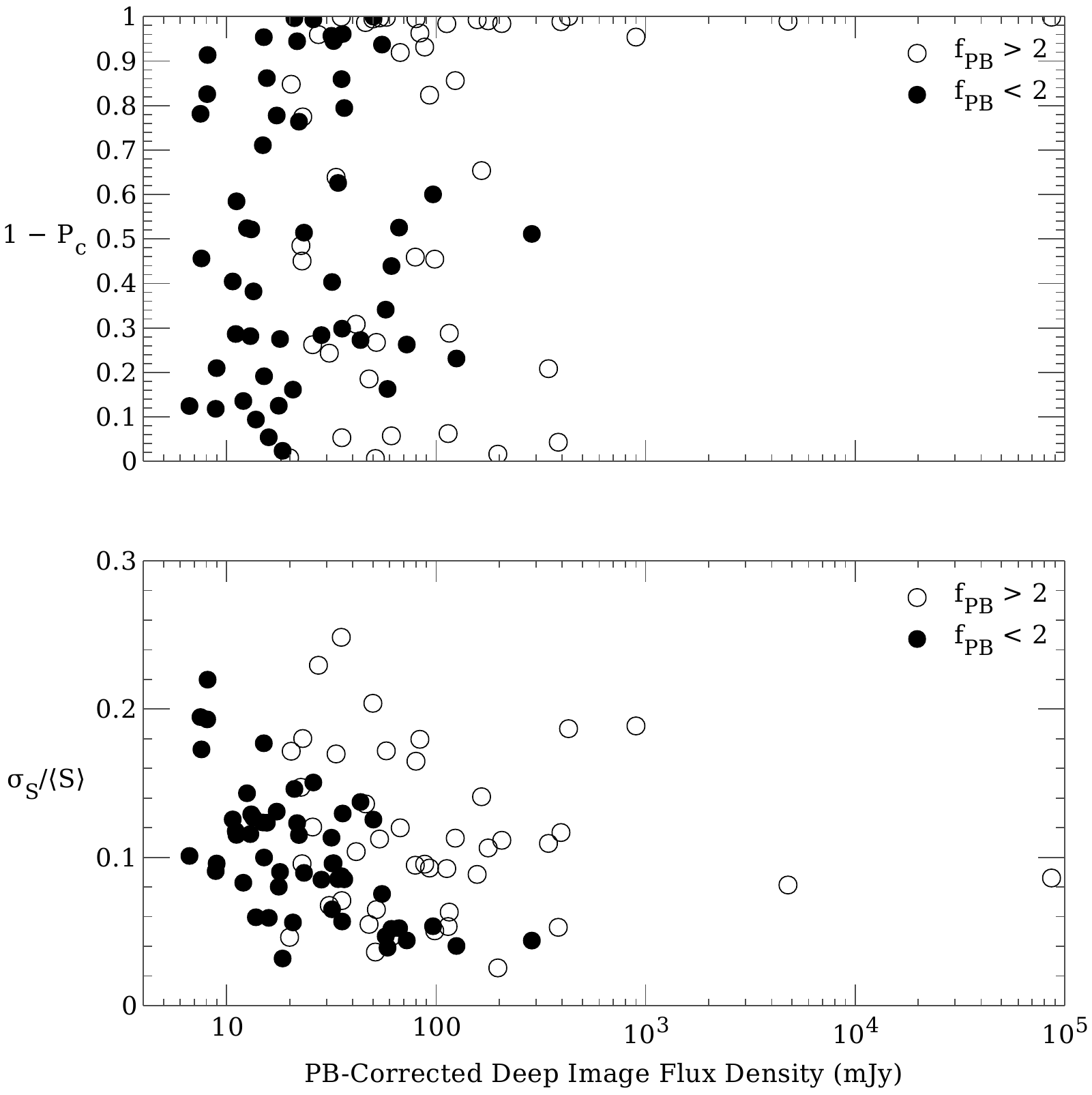}
\caption{Variability metrics as a function of deep image flux
  density. The upper panel plots $1 - P_c$ so that a higher vertical
  position indicates more variability in both panels. Among the
  sources with $f_\mathrm{PB} \le 2$, $\sigma_S / \smn$ increases for
  the less reliably-measured faint sources. The quantity $1 - P_c$
  does not show an obvious trend.}
\label{f:varvsflux}
\end{figure}

\begin{figure}[p]
\plotone{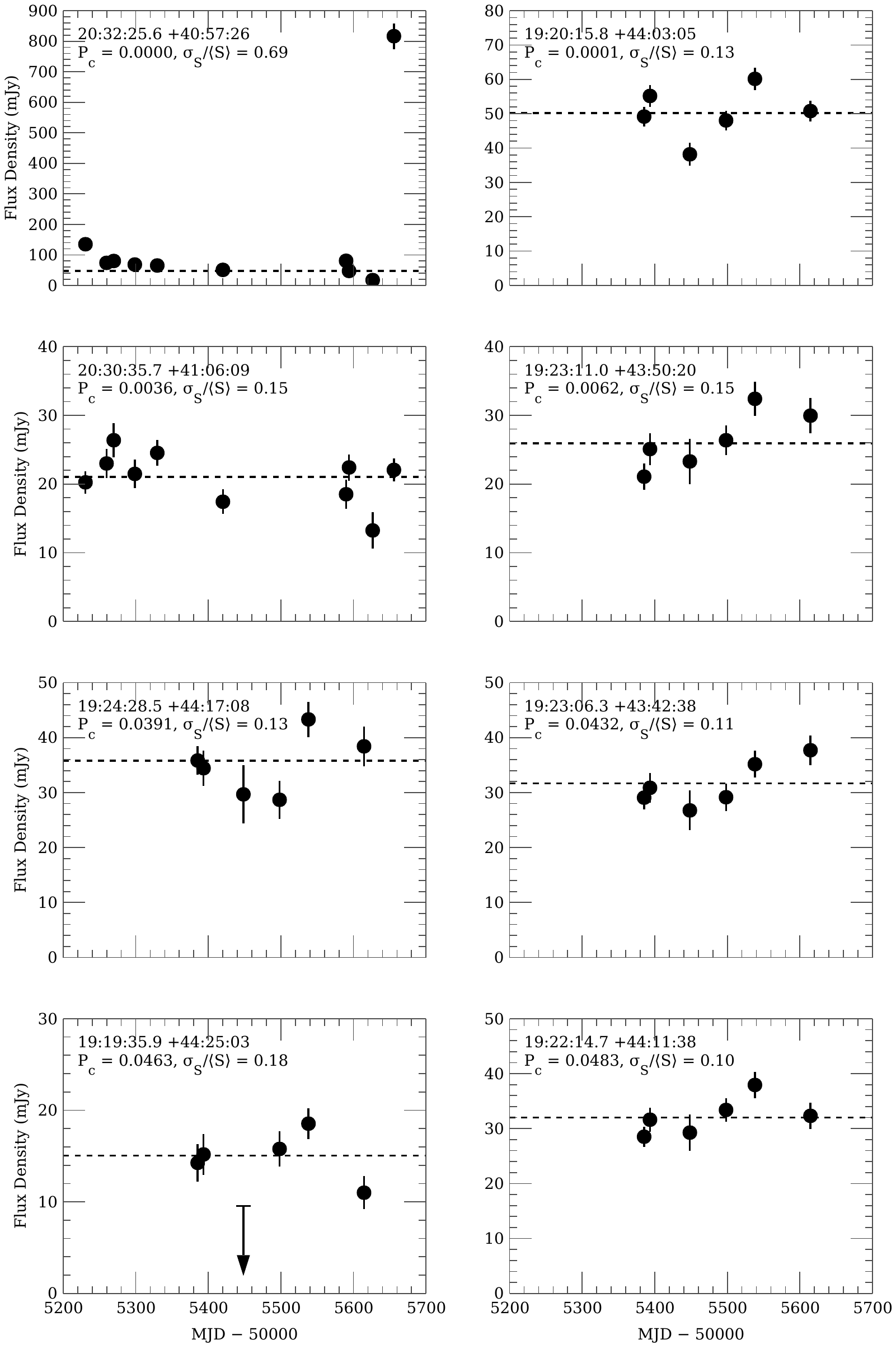}
\caption{Lightcurves of the most variable sources in this study. The
  variability metric $P_c$ increases from left to right, top to
  bottom. (Recall that lower values of $P_c$ are associated with
  higher probabilities of genuine variability.) The most-probably
  variable source, shown in the top-left panel, is \acyg, here
  identified as 20:32:25.6~+40:57:26.}
\label{f:varlc}
\end{figure}

\begin{figure}[p]
\plotone{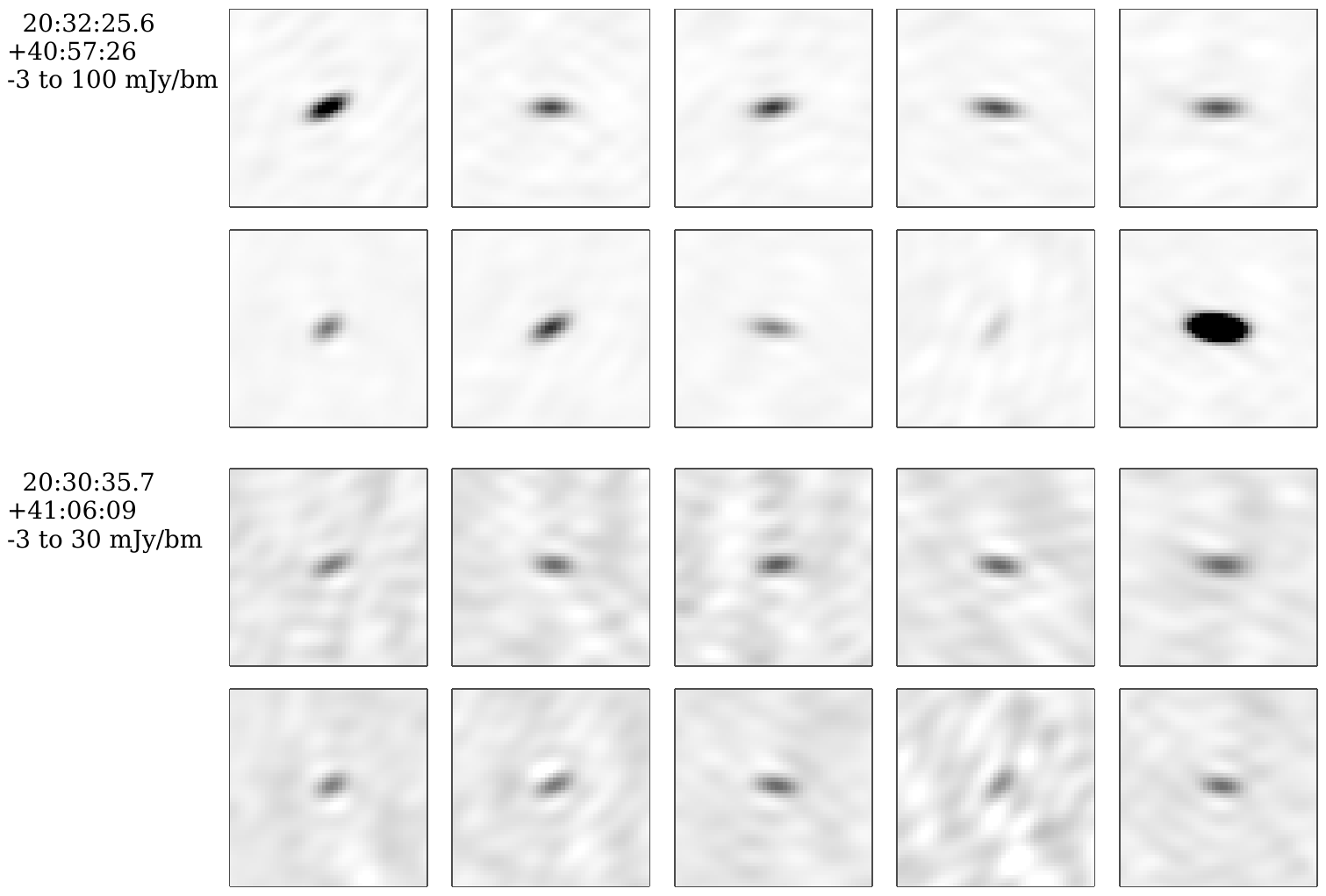}
\caption{Image cutouts (in standard J2000 RA/dec. coordinates) of the
  most variable sources in the \acyg\ pointing. Time increases left
  to right, top to bottom. The coordinates of each source and the
  bounds of the white-to-black linear intensity scale are annotated to
  the left of each set of panels. Each panel is 8.2$\times$8.2$'$. The
  top panels show \acyg\ itself, here identified as 20:32:25.6~+40:57:26.}
\label{f:stampscyg}
\end{figure}

\begin{figure}[p]
\plotone{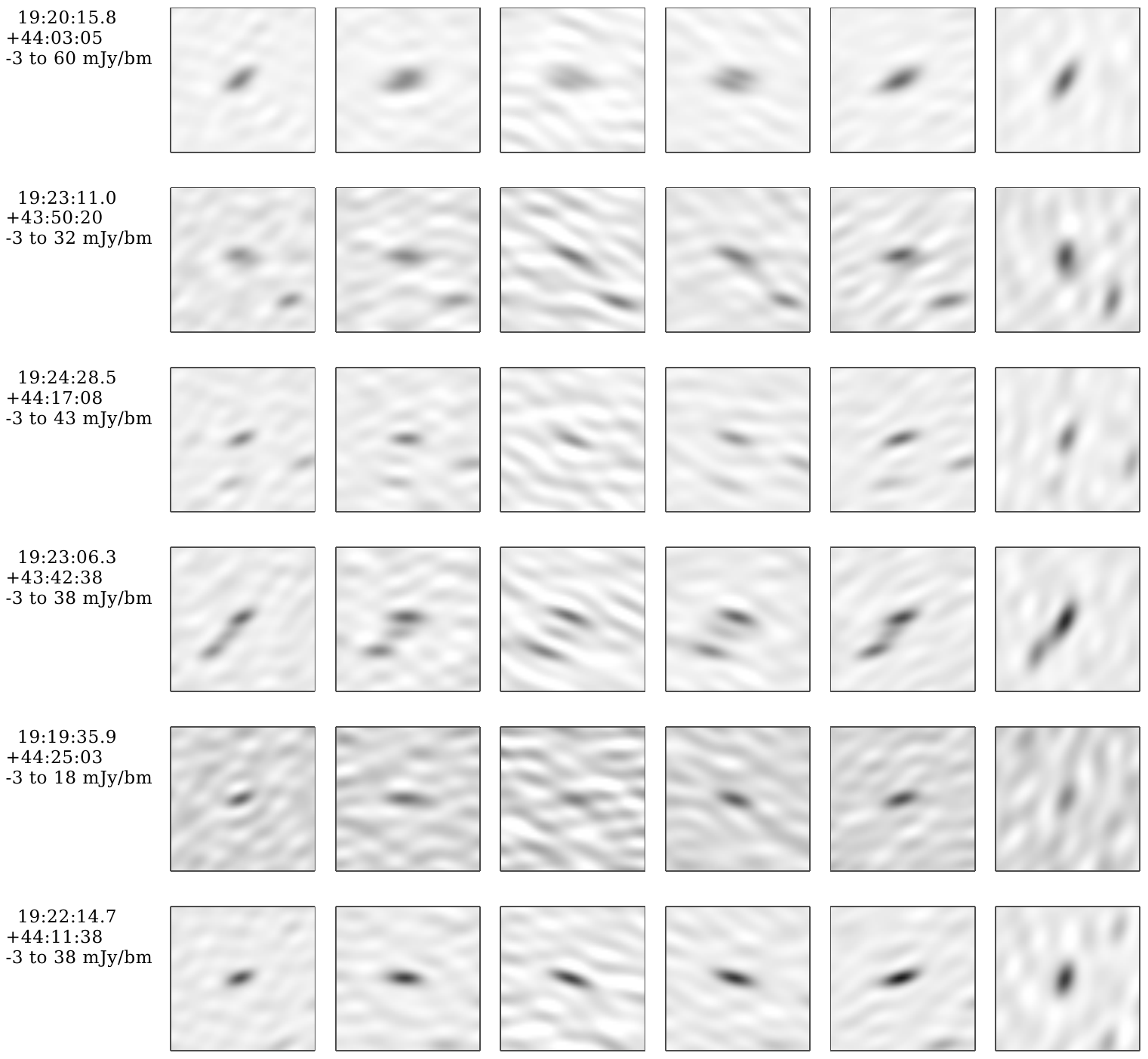}
\caption{Image cutouts of the most variable sources in the Kepler
  pointing. Layout is the same as in Figure~\ref{f:stampscyg}.}
\label{f:stampskep}
\end{figure}

Figure~\ref{f:sadlimits} shows estimates of areal densities for
various galactic radio transient phenomena and the parameter spaces
probed by several observational efforts. In this section, we describe
the different components shown in the Figure and their
derivations. This standard ``log\,$N$--log\,$S$'' plot inevitably
collapses important distinctions between different surveys and
populations, such as observing frequency, relevant timescale, and
spatial distribution. When relevant we attempt to make these
distinctions explicit in the discussion below. Most densities on the
plot are for \textit{transient} sources, as defined either empirically
(for the observations) or by an order-of-magnitude increase in flux
density (for the theoretical predictions). The exceptions are the
values shown for extreme scattering events (ESEs) and the
\citet{bhw+10} variability results, both of which correspond to \apx
50\% variability. Although we do not estimate their areal density
here, we note that flares from pre-main-sequence stars are detectable
as radio transients \citep{bpb+03,fmr08,shb08} evolving on hour
timescales at mm and cm wavelengths, and so are an additional
potential source of Galactic radio dynamism.

To provide a reference for the values we determine below, we also
include on Figure~\ref{f:sadlimits} the snapshot density of
extragalactic tidal disruption events similar to
\object{\textit{Swift}~J1644+57} as estimated by \citet{fko+12}. These
events are detected at cm wavelengths and evolve over timescales of
about a month.

\subsubsection{This Work and Forecast for Complete 3~GHz Survey}

We show the parameter space probed in this work and a forecast of the
results of a complete analysis of the 3~GHz survey data as described
in \S\ref{s:transients}. The forecast comes in the form of an areal
density upper limit should no transients be discovered. This limit
should be comparable to the extragalactic results of \pgsii\ for
sources brighter than \apx 10~mJy. Our calibrator pointings could also
be used in the transient search, but current results suggest the odds
of a successful detection would be small \citep{bfs+11,fko+12}.

\subsubsection{M-Dwarf Flares}

We show the snapshot areal density of flaring M~dwarfs as computed by
\citet{o08}, 0.11~deg$^{-2}$, taking her ``submillijansky'' detection
limit to be 0.5~mJy. The $N({>}S) \propto S^{-3/2}$ scaling here is
well-justified because M~dwarf flares are relatively faint and are
only detectable out to hundreds of pc at this limit. The rate might be
increased by similar flares from substellar dwarfs
\citep{b02}. Because coherent flares evolve quickly ($\tau \sim 60$~s)
and can be fairly narrowband ($\Delta f / f \sim 0.1$), the
detectability of these events in practice will depend strongly on
specific survey characteristics \citep{asa97}. A survey conducted in
the same manner as ASGARD would have extremely poor sensitivity to
these events, even if it appeared to reach the necessary density limit
on Figure~\ref{f:sadlimits}, because of the very different event
timescales and steep, narrow flare spectra.

\subsubsection{Active Stellar Binaries}

We again follow the analysis of \citet{o08}. \citet{dsl89} analyzed
VLA observations of 122 RS\,CVn-like active binary systems and
detected 66, finding luminosity densities ranging from \apx
$10^{14}$--$10^{18}$~erg~s$^{-1}$~Hz$^{-1}$ at 6~cm wavelengths, with
a median of \apx $10^{16}$~erg~s$^{-1}$~Hz$^{-1}$. We set a cutoff of
a factor of $>$10 luminosity increase for such a system to be
considered a transient rather than a variable. Approximately 10\% of
the systems observed had luminosity densities
${>}10^{17}$~erg~s$^{-1}$~Hz$^{-1}$. If all of these systems are
flaring median-luminosity binaries and all of the binaries have
similar variability patterns, an approximate duty cycle for active
binaries to appear as radio transients is also 10\%. We then find that
the areal density of such systems brighter than 10~mJy should be
$4\times10^{-4}$~deg$^{-2}$, taking the spatial density of active
binary systems to be \apx $6\times10^{-5}$~pc$^{-3}$ \citep{fms95}. As
with flaring M~dwarfs, these systems are only detectable out to
hundreds of pc at mJy sensitivities and thus can be treated as an
isotropic population for our purposes. Even with our somewhat generous
treatment of both the duty cycle and radio luminosity dynamic range of
these systems, they will be very difficult to detect blindly. Unlike
M~dwarf flares, however, active binary flares evolve on day timescales
and have flat, broadband spectra that are amenable to detection at cm
wavelengths \citep{og78}.

\subsubsection{X-Ray Binaries}

\citet{lrgs07} used the INTEGRAL dataset to find 74 XRBs with
$|b|<2\degr$, with 41 of those being high-mass systems. Although the
distribution of these systems is nonuniform and, in the case of
high-mass systems, appears to be linked to Galactic spiral structure
\citep{btrj12}, we compute a characteristic areal density in the GP by
assuming these sources are distributed in a region of $|\ell| <
90\degr$, $|b|<2\degr$. XRBs flare at cm wavelengths on hour-to-week
timescales, making them well-suited to the ASGARD observing strategy
\citep{hh95}. We compute typical flaring flux densities of these
systems as follows. For black hole XRBs, we use radio observations of
20 systems presented by \citet{gmf12}. The radio luminosities of
systems with multiple observations vary by factors of \apx 10, so that
marginally-detected flaring systems may appear as transients. The
maximum observed flux densities of these systems range from \apx
0.1--400~mJy, with a median of 8~mJy, assuming flat spectra and an
observing frequency of 3.09~GHz. For neutron star XRBs, \citet{mf06}
find a typical flux density of \apx 0.4~mJy and a typical radio
luminosity dynamic range of 5, so that flaring systems typically reach
\apx 2~mJy. For both classes of systems, the flaring duty cycle
appears to be \apx 1\% \citep{fkw+09,kcf05}. Combining these results,
we arrive at an areal density of \apx 10$^{-3}$~deg$^{-2}$ for flaring
XRBs and a characteristic flux density of \apx 4~mJy, where the latter
value is intermediate between the ones mentioned above with a bias
towards a smaller value to reflect the smaller luminosity dynamic
range of the NSXBs and the likelihood that the brighter flaring
systems have already been discovered. For consistency we plot this
estimate using $S^{-3/2}$ scaling but we warn that this is not as
well-motivated as in the previous cases. Although XRBs are brighter
than the aforementioned stellar systems, their rarity makes their
blind detection difficult.

\subsubsection{Extreme Scattering Events}

Most examples of interstellar scintillation affect source flux
densities by \apx 10\% \citep{lrm+08} and so would not be associated
with radio transience. Extreme scattering events, however, can cause
variations of order 50\% at the \apx 3~GHz frequencies we consider
\citep{fdjh87,fdj+94}. These events last months and so are well-suited
to detection with the ASGARD observing cadence and analysis
method. \citet{fdjh87} find an ESE duty cycle of \apx
$7\times10^{-3}$. Considering previous indications of increased
scintillation in the GP \citep{bhw+10,ofb+11} and an association
between ESEs and Galactic structure \citep{fpjd94,lfd+00}, we assume a
doubled ESE duty cycle of $1.4\times10^{-2}$ in the GP, which is
consistent with pulsar observations \citep{pk12}. The areal density of
blazars brighter than 100~mJy is \apx 0.03~deg$^{-2}$ \citep{pglp08},
and \citet{kkw+03} detected intraday variability (IDV) in 86\% (25/29)
of a blazar sample they observed. If every source subject to IDV may
experience an ESE, we find an areal density of $4\times10^{-4}$
ESE-affected sources intrinsically brighter than 100~mJy per square
degree in the GP. We note again that these sources may not be detected
as traditionally-defined transients because of both the typical scale
of the effect and the fact that ESEs involve significant dimmings, not
brightenings, of sources. The $S^{-3/2}$ scaling is of course
appropriate for extragalactic sources, but the rate of ESE incidence
may vary significantly by line of sight.

\subsubsection{50\% Variables from Becker et al. (2010)}

\citet{bhw+10} defined strong Galactic variables as sources with
$S_\textrm{max} > 1.5 S_\textrm{min}$ (i.e., $f > 1.5$;
Equation~\ref{e:modindex}) over the multi-year time baseline of the
observations. This is
approximately equivalent to $\sigma_S / \smn = 1/3$ \citep{ofb+11},
where the measurements in question are peak, not integrated, flux
densities. After correcting for an estimated extragalactic
contribution, \citet{bhw+10} find that these sources have an areal
density of 1.6~deg$^{-2}$ in their survey, not accounting for the
completeness of the underlying catalogs. We combine this normalization
with the CDF of the brightest measurements of each applicable source
and an approximation of the completeness function of the underlying
catalogs as determined by \citet{wbh05}, which we assumed applied to
both catalogs used by \citet{bhw+10}. ASGARD observations are not
directly comparable to those of \citet{bhw+10} for two major reasons:
firstly, the \kep\ pointing that we analyze is at a much higher
Galactic latitude ($b \approx +13.5$); secondly, our observations lack
the 2--15~yr time baseline over which much of the variability found by
\citet{bhw+10} occurred, although the structure function results of
\citet{of11} suggest that the level of variability on 1--10~yr
timescales is approximately constant. The difference between the
observing frequencies of the two surveys may also be relevant due to
the frequency dependence of scintillation effects in the GHz regime
\citep{hn86}. Applying the \citet{bhw+10} estimates to our sensitivity
limit and the footprint (not effective area) of the \cyg\ pointing, we
would expect to detect \apx 5 highly variable Galactic sources. In our
catalog \cyg\ is the only source with $\sigma_S / \smn > 1/3$. We
tentatively attribute this discrepancy to the comparatively short time
baseline of our study, but defer a deeper analysis until later work.

\subsubsection{GC Radio Transients from Hyman et al. (2002, 2006, 2009)}

In a total of 62 epochs of long-wavelength observations of the GC
region using the VLA and the GMRT, \citet{hlk+02,hlr+06,hwl+09}
discovered four robust GC radio transients:
\object{GCRT~J1746$-$2757}, \object{GCRT~J1745$-$3009},
\object{GCRT~J1742$-$3001}, and a radio counterpart to the X-ray
transient \object{XTE~J1748$-$288} \citep{smh+98}. Although this last
source was first detected in the X-ray band, we count it as an
independent radio transient because, based on the description of its
detection by \citet{hlk+02}, it would have been detected in a blind
search even without the X-ray detection, and it does not appear that
the X-ray detection affected the scheduling of the radio
observations. These observations were performed on a \apx monthly
cadence with several-hour integrations and so are sensitive to similar
but somewhat longer timescales than ASGARD. It is difficult, however,
to estimate an effective transient search area for this survey because
the observations were made in heterogeneous conditions and the search
methodology is not described in detail by the authors. Our best
estimate of the effective search area, attempting to take into account
the multiple overlapping pointings, different instrumental
configurations, and varying sensitivities of each observation, is
120~deg$^2$, with a characteristic detection limit of 30~mJy. The full
ASGARD dataset will probe a significantly larger area at this
limit. As with the comparison to \citet{bhw+10}, however, a direct
extrapolation of the areal density would be inadvisable:
\citet{hlk+02,hlr+06,hwl+09} observed only toward the GC itself, where
the source density is larger, and the observing wavelengths of the two
searches differ by approximately an order of magnitude.

\subsubsection{Galactic Radio Transients from Gregory \& Taylor (1986)}

\citet{ofb+11} and \citet{fko+12} have reported a density measurement
of 10$^{-3}$~deg$^{-2}$ for this survey (in their Figures~1 and 6,
respectively) but this number is erroneous (E.~Ofek, 2012, private
communication). \citet{gt86} surveyed a footprint of 500~deg$^2$ at
5~GHz over 16 epochs and discovered one bright ($>$1~Jy) transient,
\object{GT~0351+543a}, that appeared in one epoch, a phenomenology
similar to that of the events detected at Nasu Observatory
\citep{kmt+07,mdk+07,nak+07,kns+08,mnk+09}. Both day and year
timescales were sampled but intermediate ones were not; about $2/3$ of
the detected variables evolved on day timescales. The single transient
detection straightforwardly gives a density measurement of
$(1.33_{-1.30}^{+6.10})\times10^{-4}$~deg$^{-2}$ (95\% confidence
limit), but the effective search area of the survey as a function of
limiting detectable flux density is unclear due to variable survey
sensitivity. \citet{tg83} report that over the whole survey the
worst-case 3.5$\sigma$ detection limit, corresponding to an empirical
50\% survey completeness, was 17~mJy. At this flux density limit the
effective area of the survey is thus no less than 3750~deg$^2$, and a
95\% CL upper limit on the Galactic transient surface density is thus
$1.3\times10^{-3}$~deg$^{-2}$. \citet{tg83} also report a best-case
3.5$\sigma$ detection limit of 7~mJy, so that at this limit the
effective area of the survey is no more than 3750~deg$^2$, and the
Galactic transient surface density at that limit has a 95\% CL lower
bound of $1.4\times10^{-5}$~deg$^{-2}$ or \apx 0.6~sky$^{-1}$. We
advise caution in the use of these limits because it seems unlikely
that the only transient detected by \citet{gt86} would have a flux
density $>$1~Jy, when their survey should have easily been sensitive
to events $\lesssim$50 times fainter. \citet{gt86} investigated and
deemed unlikely the possibility that this event was due to
interference, however. No source was detected in later VLA followup of
the event \citep{tgdt95}.

\subsection{The Nature of the Most-Variable Sources}
\label{s:varsrcs}

We investigated the nature of the most-variable sources listed in
Table~\ref{t:vars} and plotted in
Figures~\ref{f:varlc}--\ref{f:stampskep}. Here we briefly describe our
findings, omitting the most variable source, \cyg. All of the other
sources are consistent with being AGN. The sources have values of
$\sigma_S/\smn$ ranging from 0.10 to 0.18, as compared to 0.69 for
\cyg. As mentioned in \S\ref{s:var}, none of the sources besides
\cyg\ are obviously variable upon visual inspection.

\textit{19:20:15.8~+44:03:05}: This source is a marginally-resolved
double, as confirmed by observations from the Cambridge One-Mile
telescope \citep{b78} and the VLA \citep{odo85}. Our modeling of this
source as a single component likely leads to its apparent
variability. It is a head-tail radio source found near the center of
the cluster Abell~2319. The position is consistent with that of the
infrared source \object{WISE\,J192015.63+440307.9} \citep{cwc+12},
which has colors consistent with an AGN \citep{thewise}.

\textit{20:30:35.7~+41:06:09}: This source is also detected in the
NVSS (\object{J203032+410634}) and the WSRT 327~MHz GP survey
\citep[as \object{2028+4055}]{tgc+06}. The WSRT survey included two
epochs of observations separated by several years. \citet{tgc+06}
searched for variability between the epochs, and this source was not
detected as a variable. There were no counterparts in other bands
found in standard catalogs. The flux densities at 0.327, 1.4, and 3.09~GHz are
150, 12, and 21~mJy, respectively. Although variability or an unusual
spectrum are possible explanations, we interpret the L-band
measurement as suggesting that much of this source's emission is
resolved out by the VLA.

\textit{19:23:11.0~+43:50:20}: This source is also detected in the
NVSS (\object{J192310+435016}), the Westerbork Northern Sky Survey
\citep[WENSS, as \object{B1921.6+4344};][]{thewenss}, and early
Westerbork observations of Abell clusters \citep{hm78}. It is 12\%
linearly polarized at 5~GHz \citep{bvm86} which may contribute to some
apparent variability. There were no counterparts in other bands found
in standard catalogs.

\textit{19:24:28.5~+44:17:08}: This source is also detected in the
NVSS (\object{J192428+441708}), WENSS (\object{B1922.9+4411}), and the
Green Bank 6 cm survey \citep[GB6,][as
  \object{J1924+4417}]{thegb6}. It has a generally decreasing spectrum
over this frequency range. The position is consistent with that of
\object{WISE\,J192428.35+441709.0}, which has colors consistent with
an AGN.

\textit{19:23:06.3~+43:42:38}: This source appears to be the northwest
lobe of a double source with a fainter central component. It is
detected in the NVSS (\object{J192306+434233}). The central component
is detected in the IR (\object{2MASS\,J19230940+4341460}) and X-rays
by \textit{Einstein} (\object{2E\,1921.5+4335}) and ROSAT
\citep{bsf+97}. These properties suggest that the ASGARD source in
question is a lobe of an FR\,II active radio galaxy \citep{fr74}. If
that is the case, its flux density should be virtually constant. We do not
model the faint central radio component, which will increase the
source's apparent variability.

\textit{19:19:35.9~+44:25:03}: This source is also detected in the
NVSS (\object{J191936+442503}) and possibly WENSS
(\object{B1918.0+4419}, separation \apx 20\arcsec), but not GB6. It
has a generally decreasing spectrum over the range of frequencies in
which it is detected. There were no counterparts in other bands found
in standard catalogs.

\textit{19:22:14.7~+44:11:38}: This source is also detected in the
NVSS (\object{J192214+441137}), WENSS (\object{B1920.6+4405}), GB6
(\object{J1922+4411}), and by \citet{hm78}. It has a strongly
decreasing spectrum over this range. There were no counterparts in
other bands found in standard catalogs.

\subsection{Outlook}

There is substantial evidence that the prevalence of apparent
variability increases with proximity to the Galactic plane
\citep{sfgp89,gr92,gh00,bhw+10,of11}. \citet{ofb+11} suggest that
there is also a significant increase in the typical variability
amplitude, based on the difference between their findings from
observations at $|b| \approx 7$\degr\ and those of \citet{bhw+10} at
$|b| \lesssim 1$\degr. Our observations, with $|b| \lesssim 0.5$\degr,
will allow a strong test of this conjecture if the source variability
function is flat over 1--3~yr timescales. We also probe Galactic
longitudes from $-3\degr < \ell < 7\degr$, investigating a possible
increase in radio variability with proximity to the GC as has been
tentatively measured \citep{bhw+10} and might be expected simply from
the increased source density towards \sgra\ \citep{gt87}.

As demonstrated by Figures~\ref{f:deep} and \ref{f:gpmos}, the ASGARD
dataset allows for sensitive moderate-resolution imaging of static
large-scale radio structure in the Galactic plane. Extended structures
in our deep images will include non-thermal radio filaments
\citep{yzmc84,lyzc08}, \ion{H}{2} regions \citep{bnk+03,nhr+06}, and
supernova remnants \citep{g94} at a wide range of GC separations with
implications for WIMP dark matter models motivated by \textit{Fermi}
$\gamma$-ray observations \citep{hg11,lhyz11}, the acceleration and
composition of Galactic cosmic rays \citep{poe+08,cja+11}, and the
energetics of the interstellar medium in the GC region
\citep{ms96,cjm+10}. Although there are formidable technical
challenges to full polarimetric calibration and imaging of the data,
these would provide a wealth of information on the structure of the GC
magnetic field, ionization content, and outflows \citep{nht+10,lbn11}.

\acknowledgments

We thank Peter Backus, Tom Kilsdonk, and Jon Richards for designing
and executing the SETI component of the commensal observing campaign,
which involved responsibility for essential matters such as scheduling
and frontend hardware control. We also thank Samantha Blair, Rick
Forster, Colby Guti\'errez-Kraybill, and the rest of the dedicated
HCRO staff for their tireless work to support these observations and
general ATA operations. This manuscript was improved by helpful
comments from Steve Boggs. Research with the ATA is supported by the
Paul G. Allen Family Foundation, the National Science Foundation, the
US Naval Observatory, and other public and private donors. This
research has made use of NASA's Astrophysics Data System and the
SIMBAD database, operated at CDS, Strasbourg, France. This publication
makes use of data products from the Wide-field Infrared Survey
Explorer, which is a joint project of the University of California,
Los Angeles, and the Jet Propulsion Laboratory/California Institute of
Technology, funded by the National Aeronautics and Space
Administration.

Facilities: \facility{ATA}

\bibliographystyle{yahapj}
\bibliography{pkgw}{}

\appendix
\section{Probability of Constancy From a Set of Measurements}
\label{a:pc}

In \S\ref{s:var} we consider the probability $P_c$ that a source is
unvarying, given a set of flux density measurements. In this Appendix we give
the equation used to compute this value and derive a condition that
indicates the presence of incompletely controlled systematic
measurement errors.

To compute $P_c$ we must use the full cumulative distribution function
(CDF) of the $\chi_k^2$ family. The probability of accepting the
hypothesis that a given source is constant is the probability of
finding a $\chi^2$ value at least as large as the one obtained for
that source. This is
\begin{equation}
P_c = Q(\frac{k}{2}, \frac{\chi^2}{2}),
\label{e:pc}
\end{equation}
where $Q$ is the complementary regularized $\Gamma$ function
\begin{equation}
Q(s,x) = \frac{\Gamma(s,x)}{\Gamma(s,0)},
\end{equation}
and in turn $\Gamma(s,x)$ is the upper incomplete $\Gamma$ function
\begin{equation}
\Gamma(s,x) = \int_x^\infty t^{s-1} e^{-t} \mathrm{d}t.
\end{equation}

In the theoretical case of no varying sources and truly Gaussian
errors, the observed $P_c$ will be uniformly distributed between 0 and
1. If these assumptions do not hold --- for instance, if systematics
are present --- the actual distribution of $P_c$ values can
differ. Denote the CDF of the observed $P_c$ values $F_c(p)$; that is,
$F_c(p)$ is the probability of measuring $P_c < p$ for an arbitrary
source in the ensemble. We claim that when systematics are controlled,
genuine source variability can only lead to $F_c(p) \ge p$, ignoring
variations due to finite sample size. Therefore, data that show
$F_c(p) < p$ are suggestive of uncontrolled systematics. By
construction,
\begin{align}
F_c(0) &= 0, \cr
F_c(1) &= 1,\text{ and} \cr
\mathrm{d}F_c/\mathrm{d}p &\ge 0.
\label{e:cdffirst}
\end{align}
We furthermore argue that in the ideal case the underlying $P_c$
probability density function
\begin{equation}
f_c(p) = \frac{\mathrm{d}F_c}{\mathrm{d}p}
\end{equation}
must be nonincreasing, because absent systematic measurement errors
it is impossible for sources to appear statistically less variable
(biased towards larger $P_c$) than ones that are in fact
unvarying. Therefore
\begin{equation}
\mathrm{d}^2F_c/\mathrm{d}p^2 \le 0,
\label{e:cdflast}
\end{equation}
and combining with Equations~\ref{e:cdffirst} we must in fact have
$F_c(p) \ge p$ on the unit interval. As mentioned above, the finite
number of sources being measured leads to uncertainty in $F_c$ that
allows only a probabilistic statement as to whether a given observed
$F_c$ is consistent with completely controlled systematics.

\clearpage

\clearpage

\clearpage

\clearpage

\clearpage

\clearpage

\clearpage

\clearpage

\clearpage

\clearpage

\clearpage

\clearpage

\clearpage

\clearpage

\clearpage

\clearpage

\clearpage

\clearpage

\begin{deluxetable}{llr@{.}l@{ }r@{.}lrr@{.}l}
\tablecaption{ASGARD pointing centers and 3~GHz summary
  statistics.\label{t:pointings}}
\tablehead{
  \colhead{Field} &
  \colhead{Identifier\tablenotemark{a}} &
  \multicolumn{4}{c}{Galactic Coordinates ($\ell$, $b$)} &
  \colhead{\# Epochs\tablenotemark{b}} &
  \multicolumn{2}{c}{Integ. Time (hr)\tablenotemark{b}}}
\startdata
GC & gc$-$07$+$1 & 356 & 500 & +0 & 500 & 66 & 45 & 29 \\ % gc-07+1
 & gc$-$07$-$1 & 356 & 500 & $-$0 & 500 & 0 & 0 & 00\tablenotemark{c} \\ % gc-07-1
 & gc$-$05$+$1 & 357 & 500 & +0 & 500 & 66 & 46 & 28 \\ % gc-05+1
 & gc$-$05$-$1 & 357 & 500 & $-$0 & 500 & 0 & 0 & 00\tablenotemark{c} \\ % gc-05-1
 & gc$-$03$+$1 & 358 & 500 & +0 & 500 & 64 & 14 & 20 \\ % gc-03+1
 & gc$-$03$-$1 & 358 & 500 & $-$0 & 500 & 59 & 10 & 84 \\ % gc-03-1
 & gc$-$01$+$1 & 359 & 500 & +0 & 500 & 67 & 14 & 72 \\ % gc-01+1
 & gc$-$01$-$1 & 359 & 500 & $-$0 & 500 & 76 & 22 & 44 \\ % gc-01-1
 & \sgra & 359 & 944 & $-$0 & 046 & 3 & 5 & 67 \\ % sgrastar
 & gc$+$01$+$1 & 0 & 500 & +0 & 500 & 70 & 21 & 70 \\ % gc+01+1
 & gc$+$01$-$1 & 0 & 500 & $-$0 & 500 & 5 & 0 & 80 \\ % gc+01-1
 & gc$+$03$+$1 & 1 & 500 & +0 & 500 & 65 & 16 & 87 \\ % gc+03+1
 & gc$+$03$-$1 & 1 & 500 & $-$0 & 500 & 0 & 0 & 00\tablenotemark{c} \\ % gc+03-1
 & gc$+$05$+$1 & 2 & 500 & +0 & 500 & 68 & 15 & 92 \\ % gc+05+1
 & gc$+$05$-$1 & 2 & 500 & $-$0 & 500 & 0 & 0 & 00\tablenotemark{c} \\ % gc+05-1
 & gc$+$07$+$1 & 3 & 500 & +0 & 500 & 33 & 4 & 94 \\ % gc+07+1
 & gc$+$07$-$1 & 3 & 500 & $-$0 & 500 & 67 & 8 & 45 \\ % gc+07-1
 & gc$+$09$+$1 & 4 & 500 & +0 & 500 & 70 & 19 & 73 \\ % gc+09+1
 & gc$+$09$-$1 & 4 & 500 & $-$0 & 500 & 26 & 2 & 48 \\ % gc+09-1
 & gc$+$11$+$1 & 5 & 500 & +0 & 500 & 69 & 11 & 33 \\ % gc+11+1
 & gc$+$11$-$1 & 5 & 500 & $-$0 & 500 & 3 & 6 & 12 \\ % gc+11-1
 & gc$+$13$+$1 & 6 & 500 & +0 & 500 & 72 & 16 & 55 \\ % gc+13+1
 & gc$+$13$-$1 & 6 & 500 & $-$0 & 500 & 18 & 14 & 97 \\ % gc+13-1
Cygnus & Kepler & 75 & 756 & +13 & 491 & 47 & 96 & 72 \\ % kepler
 & Cyg~X-3 & 79 & 845 & +0 & 700 & 86 & 153 & 05 \\ % cygx3
 & x3$+$0$-$2 & 79 & 845 & $-$0 & 300 & 13 & 10 & 87 \\ % x3+0-1
 & x3$+$2$+$0 & 80 & 845 & +0 & 700 & 20 & 26 & 58 \\ % x3+1+0
 & x3$+$2$-$2 & 80 & 845 & $-$0 & 300 & 10 & 3 & 96 % x3+1-1
\enddata
\tablenotetext{a}{Identifiers with numerical codes refer to offsets
  from a reference point measured in half-degree increments.}
\tablenotetext{b}{Summary statistics refer to the 3~GHz subset of data
  as described in \S\ref{s:subset}.}
\tablenotetext{c}{These pointing centers were completely observed by
  the end of the first GC season, before the switch to 3~GHz observing
  frequencies, and so have no coverage in the 3~GHz subset.}
\end{deluxetable}

\begin{deluxetable}{rlllrr}
\tablecaption{Parameters of ASGARD campaigns.\label{t:campaigninfo}}
\tablehead{
  \colhead{Subset\tablenotemark{a}} &
  \colhead{Field} &
  \colhead{Start Date} &
  \colhead{End Date} &
  \colhead{\# Epochs} &
  \colhead{Time On-Source (h)}}
\startdata
Complete & GC & 2009~May~22 & 2009~Oct~14 & 64 & 186 \\
 & GC & 2010~Mar~01 & 2010~Oct~21 & 84 & 285 \\
 & GC & 2011~Mar~01 & 2011~Apr~11 & 22 & 54 \\
 & Cygnus & 2009~Nov~24 & 2011~Apr~11 & 159 & 377 \\[3ex]
3~GHz & GC & 2010~Apr~29 & 2010~Oct~21 & 73 & 245 \\
 & GC & 2011~Mar~01 & 2011~Apr~11 & 22 & 54 \\
 & Cygnus & 2010~Feb~03 & 2011~Apr~11 & 108 & 291 \\[3ex]
This work & Cygnus & 2010~Feb~03 & 2011~Apr~11 & 30 & 83
\enddata
\tablenotetext{a}{The group of rows labeled ``Complete'' gives summary
  statistics for the complete ASGARD dataset. The group labeled
  ``3~GHz'' gives statistics for the 3~GHz subset of data as described
  in \S\ref{s:subset}. The group labeled ``This work'' gives
  statistics for the datasets analyzed in this work.}
\end{deluxetable}

\begin{deluxetable}{llr@{.}lr@{.}lr@{.}lr@{.}lr@{.}l}
\tablecaption{Eight most variable sources in the ASGARD analysis presented
  in this work.\label{t:vars}}
\tablehead{
  \colhead{R.A.} &
  \colhead{Decl.} &
  \multicolumn{2}{c}{\smn\ (mJy)} &
  \multicolumn{2}{c}{$f_\mathrm{PB}$} &
  \multicolumn{2}{c}{$P_c$} &
  \multicolumn{2}{c}{$f$} &
  \multicolumn{2}{c}{$\sigma_S / \smn$}}
\startdata
20~32~25.6 & +40~57~26\tablenotemark{a} & 47 & 5 & 1 & 0 & 0 & 0000\tablenotemark{b} & 46 & 39 & 0 & 69\\
19~20~15.8 & +44~03~05 & 50 & 2 & 1 & 1 & 0 & 0001 & 1 & 58 & 0 & 13\\
20~30~35.7 & +41~06~09 & 21 & 1 & 1 & 4 & 0 & 0036 & 1 & 99 & 0 & 15\\
19~23~11.0 & +43~50~20 & 25 & 9 & 1 & 2 & 0 & 0062 & 1 & 54 & 0 & 15\\
19~24~28.5 & +44~17~08 & 35 & 8 & 1 & 9 & 0 & 0391 & 1 & 51 & 0 & 13\\
19~23~06.3 & +43~42~38 & 31 & 7 & 1 & 4 & 0 & 0432 & 1 & 41 & 0 & 11\\
19~19~35.9 & +44~25~03 & 15 & 0 & 1 & 6 & 0 & 0463 & 1 & 69 & 0 & 18\\
19~22~14.7 & +44~11~38 & 32 & 0 & 1 & 1 & 0 & 0483 & 1 & 33 & 0 & 10
\enddata
\tablecomments{\smn\ is the weighted mean flux density of the source across
  all epochs. $f_\mathrm{PB}$ is the primary beam correction factor
  used to determine the intrinsic source flux density. $P_c$ is the
  probability that the source is constant given the measurements,
  assuming purely Gaussian errors. $f$ is the modulation index of
  \citet{bhw+10}. $\sigma_S / \smn$ is the modulation index of
  \citet{ofb+11}. See discussion in \S\ref{s:varsrcs}.}
\tablenotetext{a}{This source is \acyg.}
\tablenotetext{b}{We compute $7\times10^{-211}$ for this value, but
  this is certainly subject to numerical precision issues.}
\end{deluxetable}

\end{document}

%% file: figaliases.tex
\newcommand\figepochvsdeep{f1}
\newcommand\figintervals{f2}
\newcommand\figkepdeep{f3}
\newcommand\figdeep{f4}
\newcommand\figsingleepoch{f5}
\newcommand\figdetlimits{f6}
\newcommand\figepsrcdets{f7}
\newcommand\figlssrms{f8}
\newcommand\figgpmos{f9}
\newcommand\fignumbercounts{f10}
\newcommand\figcompleteness{f11}
\newcommand\figsadlimits{f12}
\newcommand\figvarvspbfactor{f13}
\newcommand\figpcbypbf{f14}
\newcommand\figvarvsflux{f15}
\newcommand\figvarlc{f16}
\newcommand\figstampscyg{f17}
\newcommand\figstampskep{f18}